# Colossal Magnetoresistance Manganites and Related Prototype Devices

Liu Yukuai(刘愉快), Yin Yuewei(殷月伟) [*], and Li Xiaoguang(李晓光) [*]

Hefei National Laboratory for Physical Sciences at Microscale, Department of Physics, University of Science and Technology of China, Hefei 230026, P. R. China

[*]Correspondence and requests for materials should be addressed to X. G. Li (E-mail: lixg@ustc.edu.cn) or Y. W. Yin (yyw@ustc.edu.cn)

**Abstract:**

The perovskite manganite $RE_{1-x}AE_xMnO_3$ (RE = La, Pr, Sm, etc. and AE = Ca, Sr, Ba, Pb) exhibiting colossal magnetoresistance (CMR) effect is one of strongly correlated electron systems with strong interplays among the charge, spin, orbital, and lattice degrees of freedom, such as double-exchange interaction, super-exchange interaction, Jahn-Taller-type electron-lattice distortion, and Hund's coupling, *etc*., leading to complex electronic, magnetic, and structural phase diagrams. The rich physics involved in manganites, for example, the phase separation, charge ordering, and half-metallicity, makes it as one of the hottest topics in condensed matter physics, and various prototype devices have been designed with the development of the preparation technology and scientific research. In this review paper, after a brief introduction on the most important features of crystal structure, electronic structure and phase diagram, most efforts are devoted to the following aspects: CMR effect in single phase manganite related to the field sensitive spine-charge interactions and phase separation; rectifying property and negative/positive magnetoresistance effect in manganite/Nb:STO *p-n* junctions related to the special interface electronic structure; magnetoelectric coupling effect in manganite/ferroelectric structure taking advantage of the strain, carrier density and magnetic field sensitive properties; tunneling magnetoresistance effect in tunnel junctions with dielectric, ferroelectric, and organic semiconductor spacers by using the fully spin polarized nature of manganites; and size effect on magnetic properties in manganite nanoparticles.

## 1. Introduction

The discovery of colossal magnetoresistance (CMR) effect in divalent alkaline-earth ion doped perovskite manganites $RE_{1-x}AE_xMnO_3$, where RE is trivalent rare-earth (La, Pr, Sm, etc.) and AE represents divalent alkaline-earth ions (Ca, Sr, Ba), has aroused tremendous interest due to their rich fundamental physics and great potential application in spintronics.[1-9] In fact, the magnetoresistance (MR) effect (which magnitude is typically defined as MR($H$)=$|\rho(H)-\rho(0)|/\rho(H)$ where $\rho(H)$ and $\rho(0)$ represent the resistivities with and without magnetic field) in perovskite manganites was already known early up to 1960s. For example, in 1969, a thorough MR investigation as well as the phenomenological analysis was reported for $La_{1-x}Pb_xMnO_3$ crystals.[10] However, only since mid-1990s, an enormous amount of work[5, 8] has been devoted to the manganites after the discovery of the so-called CMR in manganites (e.g., up to 127000% in $La_{2/3}Ca_{1/3}MnO_3$ thin film at 77 K and 6 T)[11]. Now, it is generally believed that the underlying origin of CMR effect is closely related to the nature of CMR manganites which are strongly correlated electron systems with interplay among the lattice, spin, charge and orbital degrees of freedom, such as double exchange interaction, Jahn-Taller effect, electronic phase separation as well as charge ordering *etc*., and has been extensively described in some review papers.[1-9]

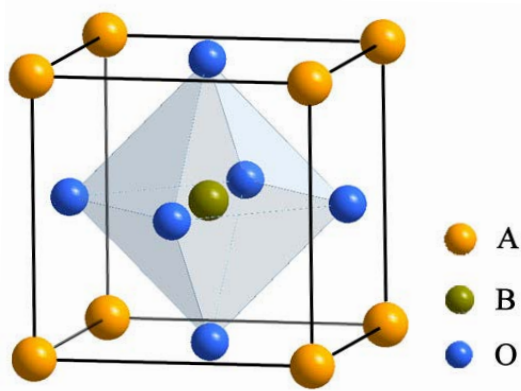

**Fig.1.** Schematic structure of the cubic perovskite $ABO_3$.

The structure of $RE_{1-x}AE_xMnO_3$ manganite is close to cubic perovskite ($ABO_3$), as schematically shown in Fig.1. The RE trivalent and AE divalent ions occupy the A-site with 12-fold oxygen coordination, while the smaller Mn ion at B site is located at the center of an oxygen octahedron with 6-fold coordination. Generally, the perovskite-based $RE_{1-x}AE_xMnO_3$ material shows lattice distortions as modifications from the cubic structure. A possible lattice deformation comes from the connective pattern of the $MnO_6$ octahedral in the perovskite structure, which is revealed by the so-called tolerance factor $t = (r_B + r_O)/\sqrt{2}(r_A + r_O)$, where $r_i$ ($i$ = A, B, and O) represents the ionic size of each element.[12]

Another lattice distortion is the deformation of the $MnO_6$ octahedral arising from the Jahn-Teller (J-T) effect, which will change the electronic structure of manganite, as shown in Fig.2.[13] Due to the crystal field, five degenerated $3d$ orbitals on the Mn-sites are subject to the partial lifting of the degeneracy into three lower-lying $t_{2g}$ orbitals and two higher-lying $e_g$ orbitals. When the occupied electrons at both orbitals are smaller than the degeneracy of the $t_{2g}$ and $e_g$ levels, the $t_{2g}$ and $e_g$ orbitals will be lifted as shown in Fig.2, namely, the so called J-T effect. In $RE_{1-x}AE_xMnO_3$ manganite, the valance states of Mn ions are of the mixture with $Mn^{3+}$ (with three $t_{2g}$ and one $e_g$ electrons) and $Mn^{4+}$ (with three $t_{2g}$ electrons only), and the proportions of Mn ions for the valence states 3+ and 4+ are $x$ and 1-$x$, respectively.

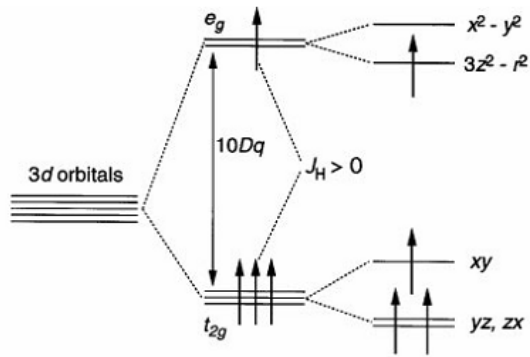

**Fig.2.** Schematic illustration of the electronic structure of $Mn^{3+}$ ions in $MnO_6$ octahedron with Jahn-Teller effect.[13]

The transfer of $e_g$ electron from $Mn^{3+}$ to $Mn^{4+}$ is the basic mechanism of electrical conduction in manganites, governed by double exchange (DE) interaction (the simultaneous jumps of the $e_g$ electron of $Mn^{3+}$ to the O $p$-orbital and the electron with same spin from the O $p$-orbital to the empty $e_g$ orbital of $Mn^{4+}$) leading to a FM state.[14] In manganites with strong DE interaction (e.g. $La_{1-x}Sr_xMnO_3$ with $x$ = 1/3), the $e_g$ electrons become delocalized for a certain doping range at low temperature, leading to a metallic behavior.[15] Because the DE interaction can be enhanced in a magnetic field by aligning the spins at adjacent Mn-O-Mn ions, the conductivity will increases, resulting in a negative magnetoresistance effect. Combined with the Hund's coupling which will remove the spin degeneracy in FM state, the separation of the spin-up ($e_g^\uparrow$, $t_{2g}^\uparrow$) and spin-down ($e_g^\downarrow$, $t_{2g}^\downarrow$) bands can be obtained as schematically shown in Fig.3.[16] One can see that the electrons almost occupy the $e_g^{1\uparrow}$ band near Fermi level, leading to a half-metallic behavior, and this fully spin-polarized is of large potential for spin electronics.[17] It is also noted that the strength of Hund's coupling can change the electronic structure, resulting a higher $t_{2g}^\downarrow/e_g^{2\uparrow}$ band with strong/weak Hund's coupling, as shown in Fig.3(a)/Fig.3(b).

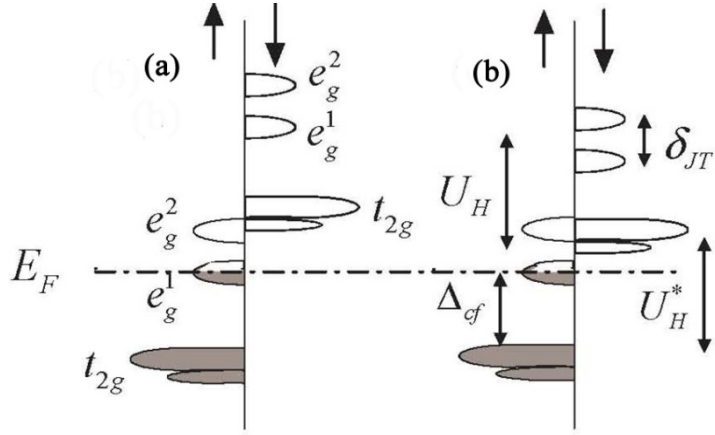

**Fig.3.** Schematic diagram of spin dependent (large arrows) density of states. (a) Strong Hund's coupling ($U_H^* \rangle \Delta_{cf} + \delta_{JT}$), and (b) weak Hund's coupling ($U_H^* \langle \Delta_{cf} + \delta_{JT}$). Here, $\Delta_{cf}$ and $\delta_{JT}$ represent the energies of the crystal field splitting and JT distortion splitting, respectively. $U_H$ or $U_H^*$ denotes the separation between $e_g^\uparrow$ and $t_{2g}^\uparrow$ or $e_g^\downarrow$ and $t_{2g}^\downarrow$ bands, and $E_F$ represents the Fermi level. [16]

Besides the DE interaction and Hund's coupling, there are also some other important factors affecting the physical properties of manganites, such as the electron-lattice interaction, FM/antiferromagnetic (AFM) super-exchange (SE) interaction between the local spins, inter-site exchange interaction between the $e_g$ orbitals (orbital ordering tendency), and intra-site and inter-site Coulomb repulsion interaction between the $e_g$ electrons *etc*.[9] These interactions/instabilities will compete with the ferromagnetic DE interaction, producing complex electronic phases.

Taking $La_{1-x}Ca_xMnO_3$ as an example, the representative electronic phase diagram is shown in Fig.4.[7] For the $La_{1-x}Ca_xMnO_3$ ($0<x<0.1$) with small amount of $Mn^{4+}$, the DE interaction is weak and the Coulomb expulsion among $Mn^{3+}$ hinders the hopping of $e_g$ electrons, leading to an insulating nature; while the super-exchange coupling (usually antiferromagnetic coupling between two next-to-nearest neighbor Mn cations through the non-magnetic O anion) between $Mn^{3+}$ causes canted antiferromagnetic (CAF).[18] For 0.1<$x$<0.2, the competition between the DE interaction and super-exchange interaction results in an insulating ferromagnetic below a certain temperature, and at further lower temperature region the charge ordered (CO) state appears.[19] The CO state with ordered $Mn^{3+}$ and $Mn^{4+}$ in real space is probably due to long-range Coulomb interaction,[20] J-T phonons interaction,[21] or super-exchange interaction,[22] but the intrinsic mechanism is still controversial.[9] With further increasing the dopant concentration $x$, the ferromagnetic metal state appears below the ferromagnetic Curie temperature ($T_C$) with a maximum $T_C$ at $x = 3/8$.[23] In this region, the system undergoes an insulator-metal transition as temperature crosses $T_C$ which moves to higher temperatures in magnetic field ($H$). For $x$>0.5, the CO transition ($T_{CO}$) is observed at temperatures higher than the $T_{CO}$ for 0.1<$x$<0.2, and accompanied with an antiferromagnetic state at low-temperature range.[24, 25] When $x$>7/8, the CO state in $La_{1-x}Ca_xMnO_3$ disappears, and the low temperature state becomes CAF again.

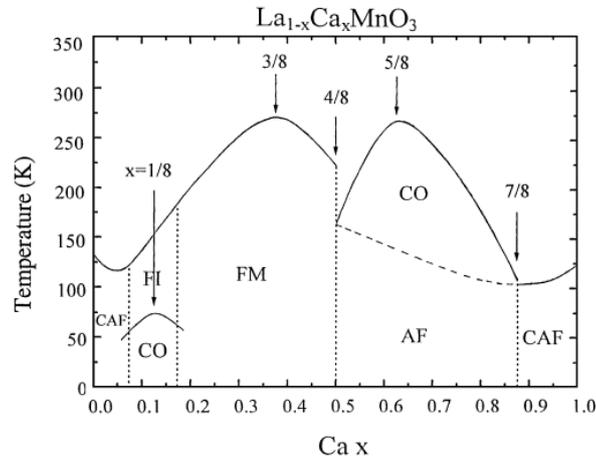

**Fig.4.** (a) Magnetic and electronic phase diagrams of $La_{1-x}Ca_xMnO_3$. The various states are: ferromagnetic insulating (FI), ferromagnetic (FM), canted antiferromagnetic (CAF), and charge-ordered (CO).[7]

It is worth mentioning that the phase boundaries are not strictly clear as these phases always coexist and compete with each other which has been experimentally observed by different characterization methods, *e.g.*, the coexistence of paramagnetic (PM) insulating and FM metallic phases in $La_{0.7}Ca_{0.3}MnO_3$ detected by scanning tunneling spectroscopy study,[26] the FM metallic and CO insulating phases in $La_{5/8-y}Pr_yCa_{3/8}MnO_3$ (0<$y$<0.4) by electron microscopy study,[27] the FM metallic and FM insulating phases in $La_{1-x}Ca_xMnO_3$ with 0.2<$x$<0.5 by $^{55}$Mn nuclear magnetic resonance (NMR) investigation,[28] and the antiferromagnetic CO structure and FM in $La_{0.35}Ca_{0.65}MnO_3$ by NMR investigation[29]. Thus, theoretical and experimental researches are rapidly converging to a unified picture pointing toward the physics of manganites in the CMR regimes clearly dominated by inhomogeneities in the form of coexisting competing phases.[9]

Therefore, because doped manganites exhibit a rich phase diagram as a function of doped concentration $x$ and the phase separation is sensitive to magnetic fields, the manganites can be used to construct a magnetoelectric coupling device as high sensitivity magnetic sensors.[30, 31] In addition, the occurrence of metallic phases with a fully spin polarized conduction band is another important promising feature in manganites for potential applications in spintronics.[17] Many prototype devices were designed by using the half-metallic property of manganites, such as magnetic tunnel junctions[32] and spin detector in spin valve devices[33].

In the present article, after a discussion on the CMR effect in single phase manganites; special attention is devoted to the newly found prototype devices based on CMR materials with heterostructure, including *p-n* junction, magnetoelectric coupling device, magnetic tunnel junction, multiferroic tunnel junction, spin-detector in spin-valve; and then, the size effect in nanosized manganites is introduced. Since the research area is broad and rapidly developing, the extensively large amount of research work based on various manganites have been published every year, so we apologize to the authors of these in advance which has unfortunately not been cited.

**2. Single phase**
**2.1 Colossal magnetoresistance and phase-separation model**

Fig.5 shows the magnetic field dependence of the resistivity for $La_{0.67}Ca_{0.33}MnO_x$ epitaxial thin film at different temperatures.[34] One can see that the film exhibits a negative ($\rho(H) < \rho(0)$) CMR effect up to 127000% at 77 K and 6 T. After annealed at 850°C/1h in a 3 atm oxygen atmosphere, Jin *et al* found that the CMR ratio of $La_{0.67}Ca_{0.33}MnO_x$ thin film can be increased from 39000% to $1.1\times10^6$% at 110 K and 6 T.[35] The CMR effect is closely related to the field sensitive complex electron-electron and electron-lattice effects in manganites which will result in tremendous conductivity variation in magnetic fields. In addition, the insulating/metallic phase separation is another field-dependent factor critical for the observed CMR effects. Both experimental and theoretical studies have shown that the typical length scale for phase-separation in manganites is from several nanometer to submicrometer scale, which depends on the material and dopant level.[27, 36-40]

Fig.6 shows the local conductivity of the $La_{0.7}Ca_{0.3}MnO_3$ single films just below $T_C$ observed by scanning tunneling spectroscopy at different magnetic fields.[26] The light colors correspond to the insulating region, whereas the dark colors represent metallic characteristics. By applying magnetic fields, a considerable fraction of insulating regions can be converted into metallic percolation path and the resistance will be suppressed accordingly leading to a significant negative magnetoresistance effect. This evolves faster in low magnetic fields ($\leq 1$ T) and continues up to 9 T. In addition, with decreasing temperature, the system becomes more metallic on average. This phase separation in a wide temperature range is observed in $La_{0.7}Ca_{0.3}MnO_3$ single films where the metallic and insulating areas are strongly temperature and magnetic field dependent.[26] The similar phase separation phenomena were also found in $Pr_{0.5}Sr_{0.5}MnO_3$,[41] $Pr_{1-x}Ca_xMnO_3$,[42] $Nd_{1-x}Sr_xMnO_3$,[43] $La_{0.45}Sr_{0.55}MnO_{3-\delta}$,[44, 45] *etc*.

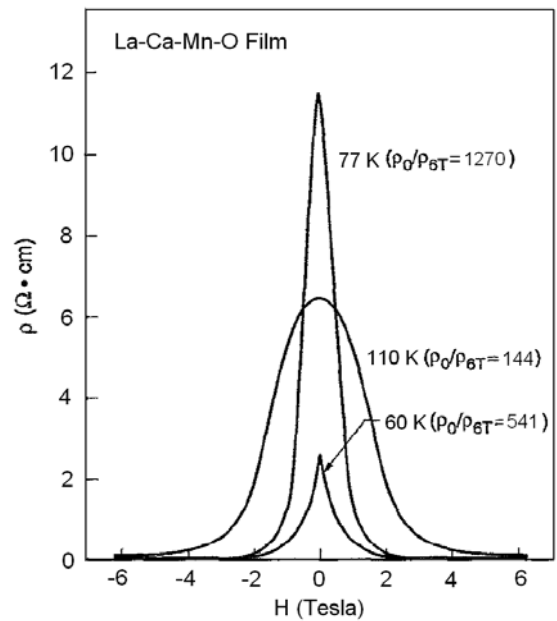

**Fig.5.** Magnetic field dependence of resistivity of $La_{0.67}Ca_{0.33}MnO_3$ thin film at different temperatures.[34]

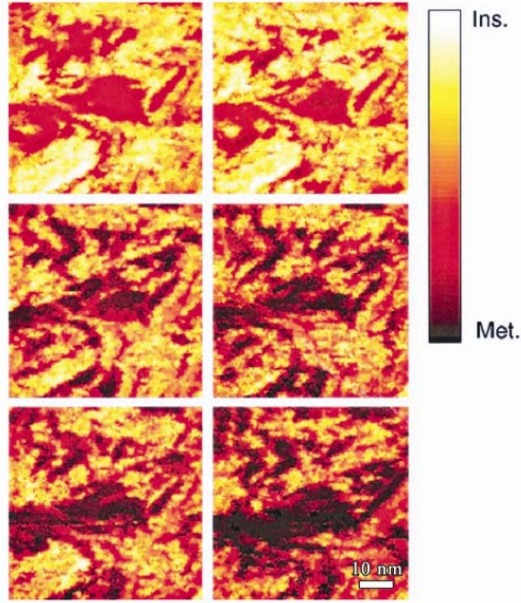

**Fig.6.** Generic spectroscopic images of the local electronic structure of $La_{0.7}Ca_{0.3}MnO_3$ single films taken just below $T_C$ in magnetic fields of 0, 0.3, 1, 3, 5 and 9 T (from left to right and top to bottom). Parts of the surface are insulating (light colors), whereas others are metallic (dark colors) or in an intermediate state.[26]

A phenomenological model was proposed to describe the relationship between the phase separation and CMR in manganites.[46, 47] For example, in a FM metallic/PM insulating phase separation system, the total resistance can be considered as a serial combination of the resistance of the coexisting phases and written as,[46]

$$R = f_{FM}R_{FM} + (1 - f_{FM})R_{PI}, \quad (1)$$

where $f_{FM}$ and $(1 - f_{FM})$ are the volume fractions, and $R_{FM}$ and $R_{PI}$ are the resistances of the FM metallic and PM insulating phases, respectively. The temperature dependence of $f_{FM}$ obeys a two energy-level Boltzmann distribution as:

$$f_{FM} = 1/[1 + \exp(\Delta U / k_B T)], \quad (2)$$

$$\Delta U = -U_0(1 - T / T_C^{mod}), \quad (3)$$

where $\Delta U$ is the energy difference between the FM metallic and PM insulating phases, $U_0$ is the energy difference between the FM metallic and PM insulating phases at $T = 0$ K, and $T_C^{mod}$ (close to $T_C$) is the fitted insulator-to-metal transition temperature used in the phenomenological model. The metallic resistance $R_{FM}$ is related to the residual resistance $R_0$, single-magnon's scattering $AT^2$ and electron-phonon interaction $BT^5$, respectively, namely, $R_{FM} = R_0 + AT^2 + BT^5$,[48, 49] while the resistance $R_{PI}$ can be described as $R_{PI} = C\exp(E_g/k_B T)$ in term of magnetic polaron picture, where constants $A$, $B$ and activation energy $E_g$ are magnetic fields dependent and become smaller due to the suppression of single-magnon scattering in FM metallic regions and the formation of magnetic polarons in PM insulating regions in magnetic fields[50]. Thus, the resistivity decreases and CMR effect appears.

Liu *et al* have employed this model to analyze the magnetic transport data of the $La_{5/8}Ca_{3/8}MnO_3$ film.[51] Fig.7(a) shows that the metal-insulator transition temperature is about $T_C = 220$ K at zero magnetic field and increases obviously with increasing magnetic fields. The temperature and magnetic field dependencies of resistances of $La_{5/8}Ca_{3/8}MnO_3$ can be well fitted (solid lines) by using Eqs.(1-3) based on the above model, as shown in Fig.7(a). The temperature dependence of ferromagnetic volume fraction $f_{FM}$ in different magnetic fields is shown in Figs.7(b) and (c), where the volume fraction $f_{FM}$ of the FM metallic phase is significantly enhanced near $T_C$ with increasing magnetic fields. It is clear that $f_{FM}$ is close to 1 or zero at temperatures well below or above $T_C^{mod}$, and increases with increasing magnetic fields especially around $T_C^{mod}$. As a result, the CMR effect manifestly appears.

The magnetic field dependencies of $T_C^{mod}$, $E_g$, and $U_0$ are shown in Fig.8. $T_C^{mod}$ increases with increasing magnetic fields as expected, see Fig.8(a). The monotonously decrease of $E_g$ in Fig.8(b) is due to the fact that the ability of the magnetic polarons to trap electrons is weakened as the spins in the polarons attempt to align along the magnetic field, which will suppress spin-disorder scattering. The field-dependent $E_g$ can be fitted (solid line of Fig.8(b)) by

$$E_g(H)/k_B = E_g(0)/k_B - \alpha H^2, \qquad (4)$$

where $E_g(0)/k_B=1002$ K and $\alpha=2.6$ K/T$^2$ are the fitting parameters. As for the $U_0(H)$, its variation with magnetic fields follows the equation

$$U_0(H)/k_B = \beta\exp(-H/H_0) + U_0(0)/k_B, \qquad (5)$$

here, $\beta$, $H_0$ and $U(0)/k_B$ are constant, as shown in Fig.8(c). With increasing magetnic fields, the increase of $f_{FM}$ and decreases of $E_g$ and $U_0$ confirm that the external magnetic fields make the PM insulating regions change to FM metallic regions. It implies that the competition between the FM metallic and PM insulating phases plays an important role in the CMR effect. It is worth mentioning that, instead of a smooth metallic-insulating phase transition around $T_C$, there is an ultra-sharp peak resistivity at metal-insulator transition temperature in La$_{5/16}$Pr$_{5/16}$Ca$_{3/8}$MnO$_3$ wires with a width comparable to the mesoscopic phase separation domains, resulting from the sudden transition between metallic and insulating states in the domain upon magnetic field and/or temperature change.[52-54]

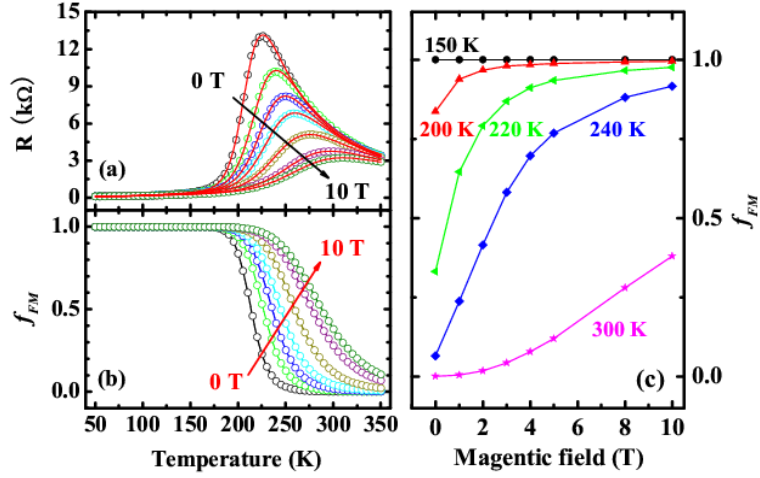

**Fig.7.** (a) Temperature dependencies of (a) resistances and (b) volume fraction of FM for the La$_{5/8}$Ca$_{3/8}$MnO$_3$ film in 0, 1, 2, 3, 5, 8 and 10 T. The solid red lines in (a) are the fitting results using Eqs.(1-3). (c) Magnetic fields dependencies of $f_{FM}$ at different temperatures.[51]

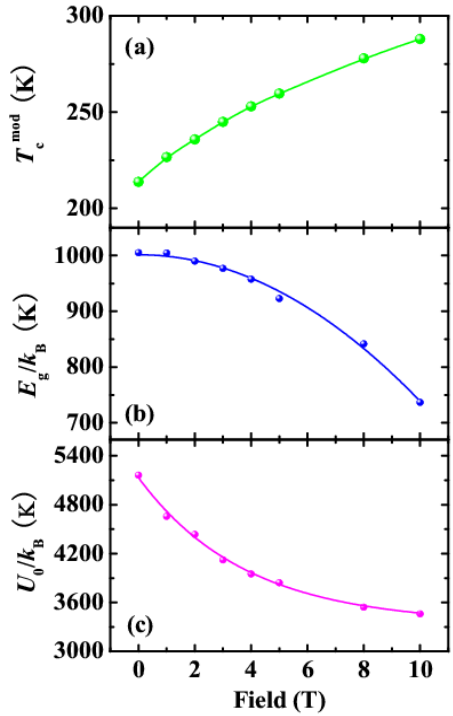

**Fig.8.** Magnetic field dependencies of $T_C^{mod}$ (a), and $E_g/k_B$ (b) and $U_0/k_B$ (c). The solid lines of (b) and (c) are plotted according to Eqs.(4) and Eqs.(5).[51]

## 2.2 Magnetic field induced melting of charge ordered states

It is well known that the $RE_{1-x}AE_xMnO_3$ manganites (e.g. $La_{1-x}Ca_xMnO_3$ with $0.5 \leq x \leq 0.9$) become insulating below the charge ordered transition temperature $T_{CO}$, because $e_g$ electrons accompanied with their J-T distortions freeze in a static order followed by a long-range antiferromagnetic spin order at Néel temperature $T_N \leq T_{CO}$. However, the real-space order of $e_g$ electrons can be destroyed by magnetic fields, it will cause an insulator-metal transition below $T_{CO}$ due to the strong spin-charge coupling, thus lead to another kind of CMR effect.[9, 55-60]

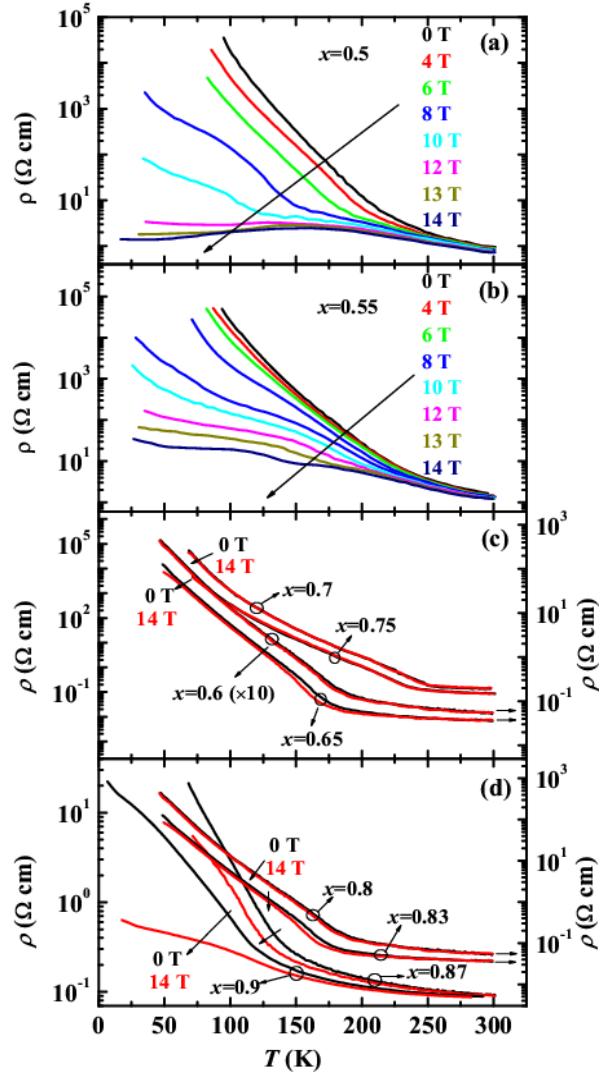

**Fig. 9.** Variations of resistivity with temperatures at different magnetic fields up to 14 T for $La_{1-x}Ca_xMnO_3$ ($x$=0.5, 0.55, 0.6, 0.65, 0.7, 0.75, 0.8, 0.83, 0.87, 0.9).[61]

Fig.9 shows the temperature dependencies of the resistivity $\rho(T)$ at different magnetic fields for $La_{1-x}Ca_xMnO_3$ ($0.5 \leq x \leq 0.9$) series.[61] With decreasing temperature, the zero-field $\rho(T)$ curve for $x$ =0.5 shows a semiconductor-like behavior, but with a discernible change in the slope near $T_{CO}$ signaling the onset of the formation of the CO state, as shown in Fig.9(a). With increasing magnetic fields from 0 T to 6 T the resistivity reduces slightly, but for 6 T $< H \leq$ 11 T the resistivity decreases dramatically although it still remains semiconductor-like. For $H >$ 12 T, a distinct change of $\rho(T)$ induced by magnetic field appears, and a metal-insulator transition occurs around 150 K, which can be viewed as a sign of the collapse of the CO state, resulting in CMR effect.[62] Similar effects were also observed for $x$ =0.55 but the material persists insulating behavior in 14 T magnetic field (Fig.9(b)). In contrast with those observed for $x$ =0.5 and 0.55, however for the samples with $x$ from 0.6 to 0.83, $\rho(T)$ and $T_{CO}$ are almost unchanged for fields up to 14 T, especially for $x$ =0.75, as shown in Figs.9(c) and (d), indicating the robustness of the ordered state near 0.75. With further increasing Ca concentration to $x >$ 0.83, $\rho(T)$ becomes weakly field dependent[63], see Fig.9(d).

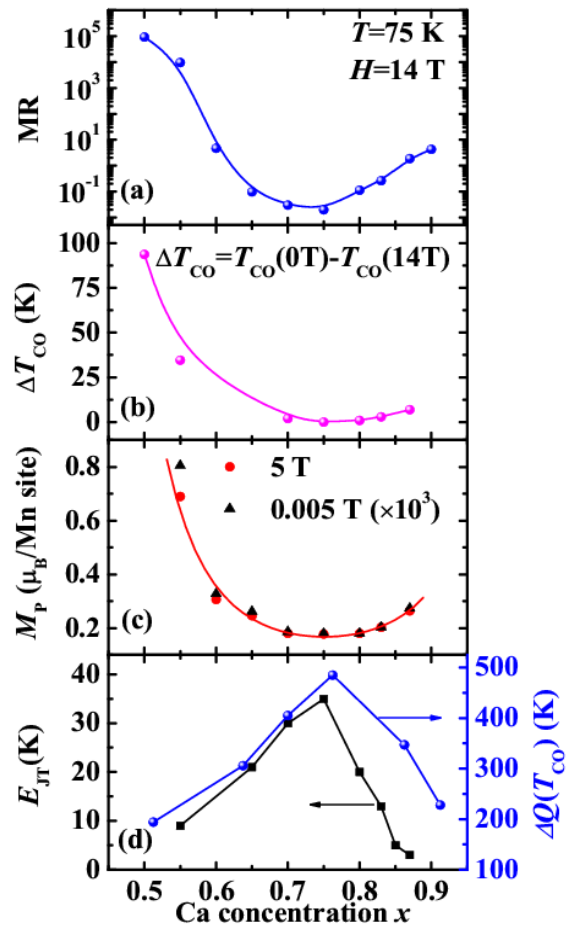

**Fig. 10.** Ca concentration $x$ dependence of (a) MR at $H$=14 T and $T$=75 K, (b) $\Delta T_{CO}$, (c) $M_P$ at $H$ = 0.005 and 5 T, (d) $E_{JT}$ and $\Delta Q(T_{CO})$ for La$_{1-x}$Ca$_x$MnO$_3$ (0.55≤$x$≤0.87).[63]

To quantitatively characterize the doping level dependence of CMR effect and the stability of CO state, Qian *et al* analyzed the MR effect at $H$=14 T and $T$=75 K and the change of $\Delta T_{CO}$ ($\Delta T_{CO}=T_{CO}$(0 T)- $T_{CO}$(14 T)) for La$_{1-x}$Ca$_x$MnO$_3$ (0.5≤$x$≤0.9), as shown in Figs.10(a) and 10(b).[63] It can be seen that the MR effect is the largest with the order of 10$^5$ for $x$=0.5, and drops rapidly with increasing $x$ from 0.5 to 0.7. Near $x$=0.75, the MR effect is minimum and CO state is most stable. For $x$>0.8, the weakened CO state becomes sensitively to external magnetic field again and the MR effect increases with increasing $x$. Further investigations of the magnetic properties, ultrasonic attenuation and specific heat of charge-ordered La$_{1-x}$Ca$_x$MnO$_3$ (0.55≤$x$≤0.87)[63] reveal that: (*i*) The peak value $M_P$ obtained from the temperature dependence of the magnetization curve at $T_{CO}$ is a minimum for $x$=0.75 sample, as shown in Fig.10(c), indicating the strength of the ferromagnetic fluctuation is the weakest near $x$=0.75 which is consistent with the variation of $\rho(T)$ in magnetic fields (Fig.9(c)). (*ii*) The J-T coupling energy $E_{JT}$ obtained from the ultrasonic attenuation measurements shows a maximum for $x$=0.75, see Fig.10(d), consistent with the change of the latent heat $\Delta Q(T_{CO})$ at $T_{CO}$, which confirms again that the CO state for $x$=0.75 is the most stable in La$_{1-x}$Ca$_x$MnO$_3$ (0.55≤$x$≤0.87).[63]

In fact, the magnetic properties for the charge ordered manganites should be also affected by phase separations[29, 64]. It is expected that the coexistence of FM and AFM phases will lead to an exchange bias (EB), as reported in La$_{1-x}$Ca$_x$MnO$_3$ (0.55≤$x$≤0.95) compound by Huang *et al*.[65] The EB effect generally manifests itself as a shift of the FM hysteresis loop from the zero-field axis due to the exchange coupling at the interface between FM and AFM spin structures.[66] Fig.11 shows the zero-field-cooled (ZFC) and field-cooled (FC, $H_{cool}$=50 kOe) magnetization-magnetic field (*M-H*) loops at 5 K for La$_{1-x}$Ca$_x$MnO$_3$ (0.55≤$x$≤0.95).[65] Fig. 11(g) shows the EB field $H_{EB}$ versus the Ca concentration $x$ at 5 K. The $H_{EB}$ is defined as $H_{EB} = |H_1 + H_2|/2$, where $H_1$ and $H_2$ denote the left and right coercivity fields, respectively. The EB effect is clearly related to the FM phases induced by the electronic phase separation in the AFM background for the charge ordered manganites. This intrinsic inhomogeneity induced EB effect is important for designing related devices.[65]

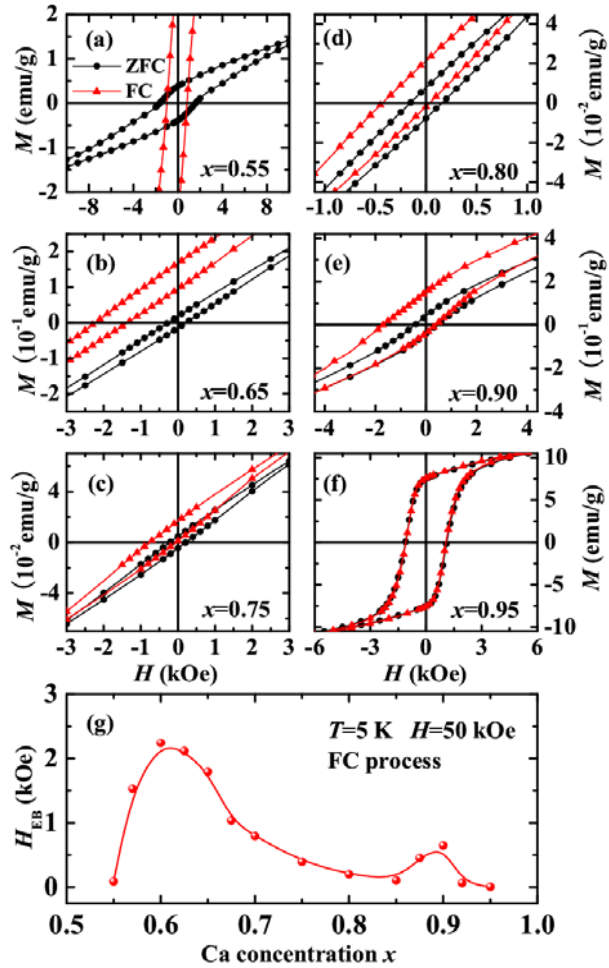

**Fig.11.** (a-f) Typical ZFC and FC *M-H* loops of the $La_{1-x}Ca_xMnO_3$ samples with $0.55 \leq x \leq 0.95$ at 5 K, and (g) Ca concentration *x* dependences of $H_{EB}$ at 5 K. Both the measuring and FC fields ($H_{cool}$) are 50 kOe.[65]

## 3. Heterostructure devices

The colossal magnetoresistance effect of perovskite manganites in single phase is one of the outstanding phenomena observed in transition-metal oxides and has stimulated intensive studies for many interesting physical properties associated with competing degrees of freedom of spin, charge, orbital and lattice.[5] Therefore, various devices were designed by using the rich physics involved in manganites, including the hole doped *p*-type semiconductivity, complex electronic structure, phase separation, half metallicity etc.[5] Several key prototype devices will be introduced as follows.

### 3.1 *p-n* junction

As we know, the low level doped $La_{1-x}AE_xMnO_3$ (AE=Ca, Sr and Ba) manganites always exhibit *p*-type semiconductivity, which can be used to construct novel *p-n* junctions with *n*-type Nb-doped SrTiO$_3$ (SNTO) substrates. The typical temperature dependent current-voltage (*I-V*) curves for manganite/SNTO *p-n* junctions, taking La$_{0.67}$Sr$_{0.33}$MnO$_3$/SNTO as an example, is shown in Fig.12, presenting a good rectifying behavior where the junction resistance reduces significantly with increasing bias. It should be mentioned that these manganite/titanate *p-n* junctions exhibit magnetically and electrically tunable physical properties due to the rich band structure at the interface of the heterostructures.[16, 67-71]

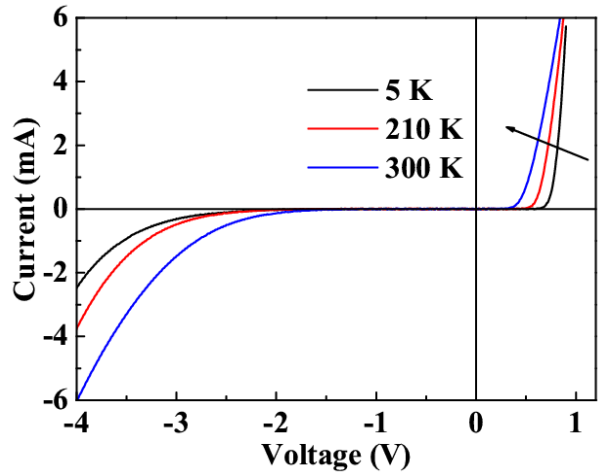

**Fig.12.** *I-V* curves of La$_{067}$Sr$_{0.33}$MnO$_3$/SNTO *p-n* junction at 5, 210, and 300 K.

### 3.1.1 Electric field manipulating metal-insulator transition

Taking advantage of "band gap engineering", the manganite/titanate *p-n* junctions can offer large changes in both electric and magnetic properties.[67, 72] In La$_{0.9}$Ba$_{0.1}$MnO$_3$/SNTO *p-n* junction with Nb-0.01 wt% doped SrTiO$_3$, besides the good rectifying effect, the junction property strongly depends on the bias voltage $V_{bias}$, as shown in Fig.13.[67] At low bias $V_{bias}$ from +0.6 to +0.8 V, the *p-n* junction is of semiconductivity and the junction resistance monotonously increases with decreasing temperature. However, with further increasing bias, a semiconductor-metal trasnsition ($T_p$) in junction resistance-temperature curves appears at $V_{bias}$=+1.0 V, and it increases gradually with increasing $V_{bias}$. Furthermore, this giant electric field dependent junction resistances are different for junctions with different thicknesses (15 nm, 30 nm) of the La$_{0.9}$Ba$_{0.1}$MnO$_3$ as shown in Fig.13(a),[67] which is due to the more significant manipulation of carrier density in a thinner La$_{0.9}$Ba$_{0.1}$MnO$_3$ film by the carrier injection from *n*-type materials. Fig.13(b) shows the temperature dependence of magnetoresistance in a magnetic field of 5 T for La$_{0.9}$Ba$_{0.1}$MnO$_3$/SNTO with 30 nm

La$_{0.9}$Ba$_{0.1}$MnO$_3$ at different bias voltages.[67] The junction resistance was reduced by magnetic field in all temperature ranges, and the MR [($R_{junction}$($H$=0 T)-$R_{junction}$($H$=5 T)]/$R_{junction}$($H$=0 T)×100%) increases monotonically with decreasing temperature and bias. The semiconductor-metal trasnition and magnetoresistance manipulated by low bias voltage in manganite/titanate $p$-$n$ junction indicate the double exchange ferromagnetism in manganite can be switched by electric fields.[67]

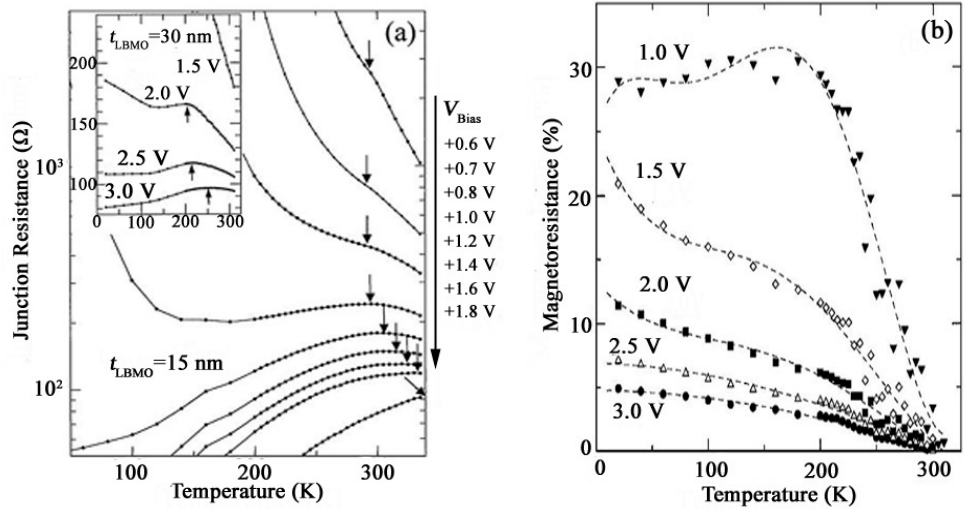

**Fig.13.** (a) Temperature dependences of junction resistances of a La$_{0.9}$Ba$_{0.1}$MnO$_3$(15 nm)/SNTO heterojunction measured under various bias voltages. The arrows indicate metal-insulator transition temperatures. The inset shows that of $t_{LBMO}$=30 nm. (b) Temperature dependence of magnetoresistance ($H$=5 T) measured under various bias voltages of the La$_{0.9}$Ba$_{0.1}$MnO$_3$/SNTO heterojunction ($t_{LBMO}$=30 nm). The MR ratio is defined as ($R_{junction}$($H$=0 T)-$R_{junction}$($H$=5 T))/ $R_{junction}$($H$=0 T) ×100%.[67]

### 3.1.2 Positive magnetoresistance

It is interesting that both negative MR and positive MR effects can be observed in the manganite/titanate junctions.[16, 70, 71, 73, 74] Fig.14 shows magnetic field dependencies of MR for La$_{0.82}$Ca$_{0.18}$MnO$_3$/SNTO $p$-$n$ junction at 60, 90, 180, and 240 K. It can be seen that the MR is always negative within the whole measured temperature range at a small bias current ($I$ =3 μA), and it is still negative at 60 and 90 K at a larger bias current ($I$ = 300 μA) but becomes positive at 180 and 240 K. While only negative MR is observed for the La$_{0.7}$Ca$_{0.3}$MnO$_3$/SNTO $p$-$n$ junction.[16, 75] Similar manipulations of MR from negative to positive were also reported in La$_{0.7}$Ce$_{0.3}$MnO$_3$/SNTO $p$-$n$ junction,[71] La$_{0.32}$Pr$_{0.35}$Sr$_{0.33}$MnO$_3$/SNTO $p$-$n$ junction,[73] SrTiO$_{3-\delta}$/La$_{0.9}$Sr$_{0.1}$MnO$_3$/SrTiO$_{3-\delta}$/La$_{0.9}$Sr$_{0.1}$MnO$_3$ ($n$-$p$-$n$-$p$) heterojunctions,[76] *etc.*

The results mentioned above indicate that the negative and positive MR in manganite/SNTO depend on not only temperature and bias current, but also the carrier concentration, which is directly related to the band structure at the *p-n* interface.[16, 75]

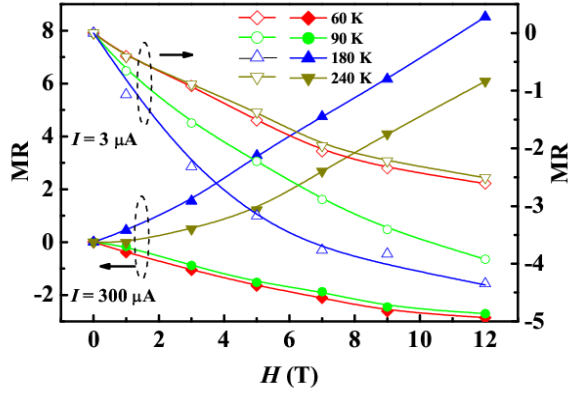

**Fig.14.** Magnetic field dependence of MR at 60, 90, 180, and 240 K for two different bias currents $I$ =3 μA (open symbols) and 300 μA (closed symbols) for La$_{0.82}$Ca$_{0.18}$MnO$_3$/SNTO *p-n* junction. [16]

As discussed in Fig.3, the Fermi surface lies in $e_g^{1\uparrow}$ band, and the $e_g^{2\uparrow}$ and $t_{2g}^{\downarrow}$ bands are above and close to the Fermi level. Therefore, with increasing the electric field applied across the *p-n* junction, the injected electrons from SNTO to manganite will occupy these empty bands from low to high. When the injection electrons from SNTO to La$_{0.82}$Ca$_{0.18}$MnO$_3$ predominantly occupy the $e_g^{1\uparrow}$ ($e_g^{2\uparrow}$) band, the injected carriers are of majority spin direction (up) and the junction persists the negative MR. However, when the electrons from SNTO occupy the $t_{2g}^{\downarrow}$ band, because the spin orientation of $t_{2g}^{\downarrow}$ is antiparallel to the majority spin direction, therefore, increasing the magnetic field will increase the spin scattering in the $t_{2g}^{\downarrow}$ band, resulting in a positive MR.[16, 75]

By selecting proper dopant composition in manganites, one can achieve a large MR effect in manganite/SNTO *p-n* heterojunctions at low magnetic field and high temperature, which is more suitable for spintronic applications than CMR effect in single phase at high magnetic field. Fig.15 shows the negative bias voltage dependence of the positive MR (MR=$R(H)/R(0)-1$) in La$_{0.9}$Sr$_{0.1}$MnO$_3$/SNTO *p-n* heterojunction.[70] In 5 Oe, a positive MR value of 10.6% at 290 K can be obtained.

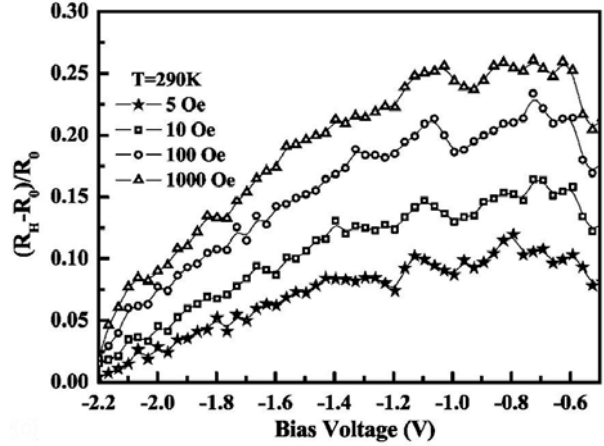

**Fig.15.** The variation of MR values of $La_{0.9}Sr_{0.1}MnO_3$/SNTO *p-n* heterojunction with the negative bias voltage at different magnetic fields at 290 K.[70]

## 3.2 Magnetoelectric coupling devices
### 3.2.1 Ferroelectric controlled electric and magnetic properties of manganites

Due to the sensitivity of the spin, charge, orbital and lattice degree of freedoms to external stimulus in manganites, the magnetic and electric properties of manganites can be tuned in situ by using a neighbored ferroelectric through strain and/or charge manipulation in the manganites/ferroelectric multiferroic heterostructures.[31, 51, 77, 78] The ferroelectric materials $(1-x)Pb(Mg_{1/3}Nb_{2/3})O_3$-$xPbTiO_3$ (PMN-PT) with composition near the morphotropic phase boundary ($x$=0.33) exhibit outstanding piezoelectric activities and giant remanent polarization, and can be used as ferroelectric substrates to grow the manganite thin film.[77, 79-85] Figs.16(a) and 16(b) show the typical butterfly-like strain-electric field hysteresis loops and rectangular-like polarization-electric field loops of PMN-PT single crystals. Figs.16(c) and 16(d) show the strain and ferroelectric field effects on the resistances of the $Pr_{0.5}Ca_{0.5}MnO_3$ film in $Pr_{0.5}Ca_{0.5}MnO_3$/PMN-PT heterostructures at different temperatures.[86] A roughly symmetrical butterfly-like resistance-electric field (*R-E*) hysteresis loop at 296 K is a typical behavior of the resistance modification due to the polarization-rotation-induced strain in the PMN-PT substrate. The strain-induced resistance manipulation implies that the intrinsic double-exchange interaction is strongly influenced by the in-plane strain.[87] With decreasing temperature to 180 K (Fig.16(d)), the rectangular-like *R-E* loop appears, indicating an enhanced ferroelectric field effect with decreasing temperature in $Pr_{0.5}Ca_{0.5}MnO_3$/PMN-PT heterostructure. Thus one can control the properties of manganites by electric voltage in manganite/ferroelectric heterojunctions, due to the transferred electric-field-induced strain to the manganite and modified charge carrier density of the manganite by ferroelectric fields.

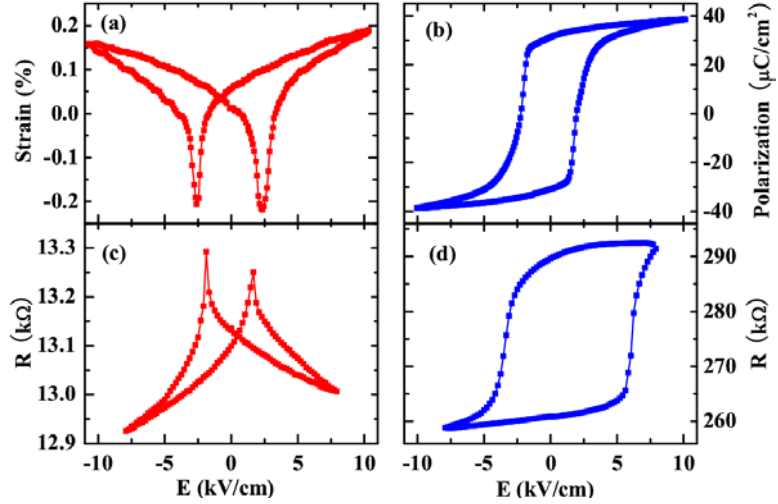

**Fig.16.** Typical (a) butterfly-like strain-electric field and (b) rectangular-like polarization-electric field hysteresis loops of PMN-PT single crystals at room temperature. Changes in the resistances of the $Pr_{0.5}Ca_{0.5}MnO_3$ film at (c) 296 K and (d) 180 K as a function of bipolar electric field applied to the substrate.[86]

Recently, Yang *et al* reported the anisotropic magnetic properties controlled by anisotropic strain effects.[88] They found that the remnant magnetization of $La_{2/3}Sr_{1/3}MnO_3$ along [100] direction was suppressed through strain effect, while this along [01$\bar{1}$] was enhanced for film deposited on (011) PMN-PT single substrate. The tunabilities of the remnant magnetizations along the [100] and [01$\bar{1}$] directions are about -17.9% and +157% in electric field of +7.27 kV/cm, respectively.[88] This large anisotropic remnant magnetization tunability may find potential applications in the electrically written and magnetically read memories.[89] Likewise, the valence state of Mn induced by electronic charge manipulation through field effect was observed in $Pb(Zr_{0.2}Ti_{0.8})O_3/La_{0.8}Sr_{0.2}MnO_3$ multiferroic heterostructure, resulting in the switching of magnetism in $La_{0.8}Sr_{0.2}MnO_3$ with the electric field "on" and "off", giving rise a large magnetoelectric coupling effect.[90, 91]

Besides the strain and field effects, Yu *et al* found that the electronic orbital reconstruction at the manganite/ferroelectric interface will directly influence the coupling between $La_{0.7}Sr_{0.3}MnO_3$ and $BiFeO_3$ and result in the pinned uncompensated spins required for exchange bias (EB).[92] This EB can be controlled by electric fields, where two distinct exchange-bias states can be reversibly switched by the ferroelectric polarization of BiFeO$_3$.[93]

**3.2.2 Magnetic field effects on the magnetoelectric coupling in manganite/ferroelectric heterostructures**

In fact, the magnetoelectric coupling effect can be impacted by both electric and magnetic fields.[30, 94] A colossal magnetocapacitance (MC=[$\varepsilon(H)$-$\varepsilon(0)$]/$\varepsilon(0)$, where $\varepsilon(H)$ and $\varepsilon(0)$ are the dielectric constants with and without magnetic fields) up to 1100% near the Curie temperature $T_C$ = 220 K of La$_{5/8}$Ca$_{3/8}$MnO$_3$ in BiFeO$_3$/La$_{5/8}$Ca$_{3/8}$MnO$_3$ heterostructure is observed, see Fig.17,[78] which is three orders of magnitude higher than that reported in a single phase BiFeO$_3$ system.[95, 96] The colossal magnetocapacitance effect can be achieved in a wide frequency region from 200 kHz to 2 MHz and a wide temperature range from 190 K to 260 K.

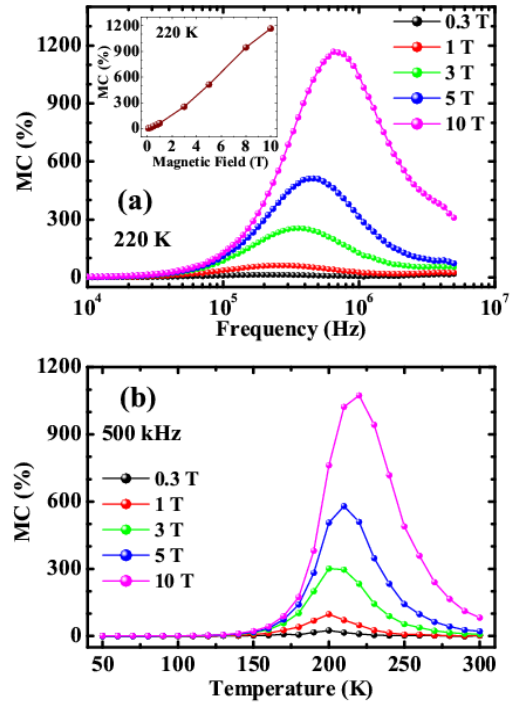

**Fig.17.** Frequency dependencies of the MC effect at 220 K. The inset shows the magnetic field dependence of maximum MC value at 220 K. (b) Temperature dependencies of the MC effect at a frequency of 500 kHz in magnetic fields up to 10 T.[78]

Fig.18 shows the temperature and magnetic field dependencies of the remanent ferroelectric hysteresis loops for epitaxial BiFeO$_3$/La$_{5/8}$Ca$_{3/8}$MnO$_3$ film.[51] With increasing magnetic fields, the room-temperature polarization has little response to magnetic fields (Fig.18(a)), and also no obvious differences in the *P-E* hysteresis loops are observed at temperatures far below the ferromagnetic transition temperature $T_C \sim 220$ K of $La_{5/8}Ca_{3/8}MnO_3$ (Figs.18(c) and 18(d)). However, in the temperature range near $T_C$, the values of the apparent coercive electric field $E_c$ gradually decrease with increasing magnetic fields, as shown in Fig.18(b). Based on the phase-separation model discussed in section 2.1,[46] it is found that $E_c$ decreases rapidly with increasing the ferromagnetic volume fractions $f_{FM}$ of $La_{5/8}Ca_{3/8}MnO_3$ especially as $f_{FM}$ closes to 1 near 220 K, as shown in Fig.19,[51] which reflects the effect of $La_{5/8}Ca_{3/8}MnO_3$ phase separation on the magnetoelectric coupling of the interface between $BiFeO_3$ and $La_{5/8}Ca_{3/8}MnO_3$. Through analyzing the frequency dependencies of the impedance module |Z| of $BiFeO_3/La_{5/8}Ca_{3/8}MnO_3$ heterostructure, $BiFeO_3$ and $La_{5/8}Ca_{3/8}MnO_3$ in different temperatures and magnetic fields, Liu *et al* found that the obvious decrease of the apparent coercive field $E_c$ with increasing magnetic fields around $T_C$ is closely related to the voltage drops across the $La_{5/8}Ca_{3/8}MnO_3$ layer and the magnetoelectric-coupled interface between $BiFeO_3$ and $La_{5/8}Ca_{3/8}MnO_3$, which can be influenced by magnetic fields due to the phase separation in $La_{5/8}Ca_{3/8}MnO_3$.[51]

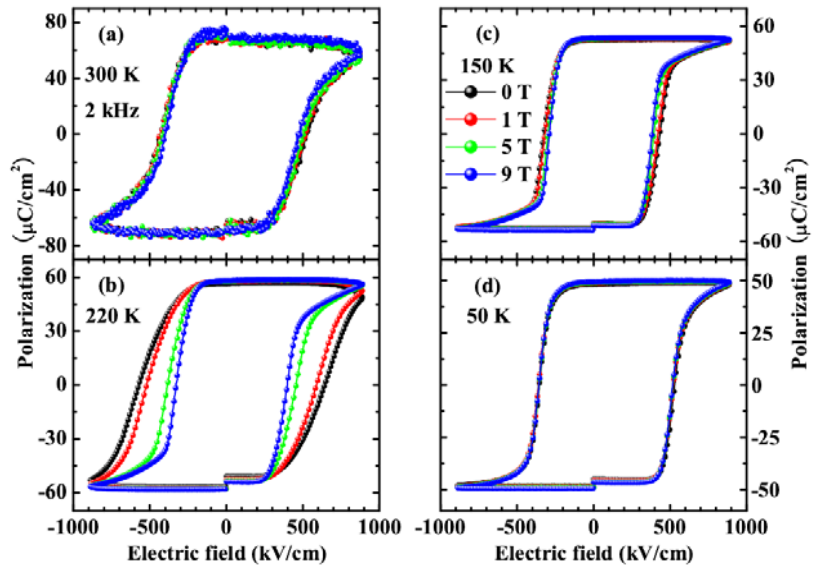

**Fig.18.** Remanent ferroelectric hysteresis loops of $BiFeO_3/La_{5/8}Ca_{3/8}MnO_3$ at $f = 2$ kHz in magnetic fields of 0, 1, 5 and 9 T at different temperatures: (a) $T = 300$ K, (b) $T = 220$ K, (c) $T = 150$ K and (d) $T = 50$ K. [51]

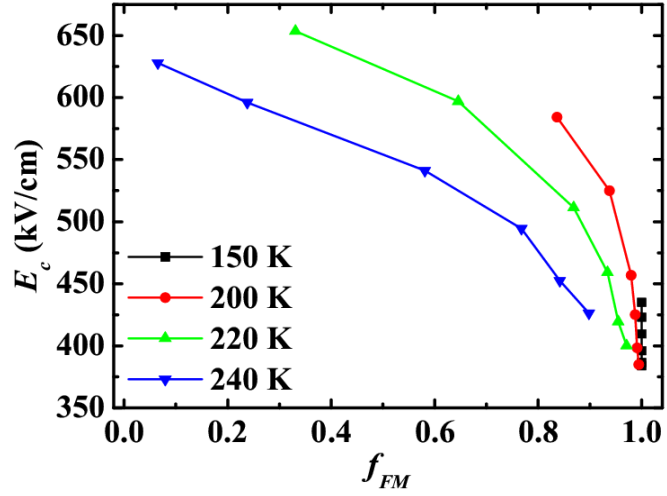

**Fig.19.** FM volume fraction $f_{FM}$ dependencies of $E_c$ in $La_{5/8}Ca_{3/8}MnO_3$.[51]

## 3.3 Spintronic devices based on half-metallic property in manganites
### 3.3.1. High TMR effect in Magnetic tunnel junction

The CMR effect in both bulk and single film always needs a large magnetic field, which limits its application in spintronic device.[8] Therefore, alternative devices were designed to overcome this obstacle, such as magnetic tunnel junction (MTJ) which takes advantage of the high spin polarization in FM manganite below Curie temperature.[17, 97-99] A magnetic tunnel junction is composed of two ferromagnetic layers (CMR materials) separated by an ultrathin insulating barrier.[17, 97, 100-102] Because these tunnel junction devices allow a true on-off operation, they are very suitable for the nonvolatile memory devices, such as magnetic random access memories (MRAM). This device was first realized in $La_{0.67}Sr_{0.33}MnO_3/SrTiO_3/La_{0.67}Sr_{0.33}MnO_3$ (LSMO/STO/LSMO) junction with tunneling magnetoresistance (TMR) of 86% at 4.6 K.[32] Obata *et al* have fabricated LSMO/STO/LSMO junctions with a thin STO tunnel barrier,[100] they found the TMR disappears above 200 K for a junction with a thickness of STO $t_{STO}$=2.4 nm, whereas for a thinner junction ($t_{STO}$=1.6 nm) the TMR value was still observed up to 270 K. Furthermore, extremely large TMR response up to 1800% was obtained in LSMO/STO/LSMO/Co/Au device, as shown in Fig.20, where the spin-polarization of LSMO at the interface with STO can be reach at least 95%.[97] However, when the spin-polarizations of the two ferromagnetic electrodes are opposite sign, an inverse TMR is observed with a smaller resistance in the antiparallel configuration, such as in LSMO/I/Co junctions where I is an insulating barrier of STO, $Al_2O_3$, or $Ce_{0.69}La_{0.31}O_{1.845}$.[102] An unexpected result is the fact that the sign of TMR depends on the nature of the insulating barrier: a maximum tunneling current is observed in parallel configuration with $Al_2O_3$ where an inverse TMR corresponding to a maximum tunneling current for antiparallel configuration is observed with STO.[102] Interestingly, in all-manganite tunnel junction with $La_{0.7}Ca_{0.3}MnO_3$ as ferromagnetic electrodes and $La_{0.3}Ca_{0.7}MnO_3$ with a charge-ordered insulator below 260 K as a tunnel barrier, a large TMR up to 500% at 6 K was observed, where it is found that the response of this junction is not only determined by the high spin-polarization of FM electrodes but also by novel interface-induced phases.[103] These results confirm that the metal-oxide interfaces play an important role in MTJ devices.

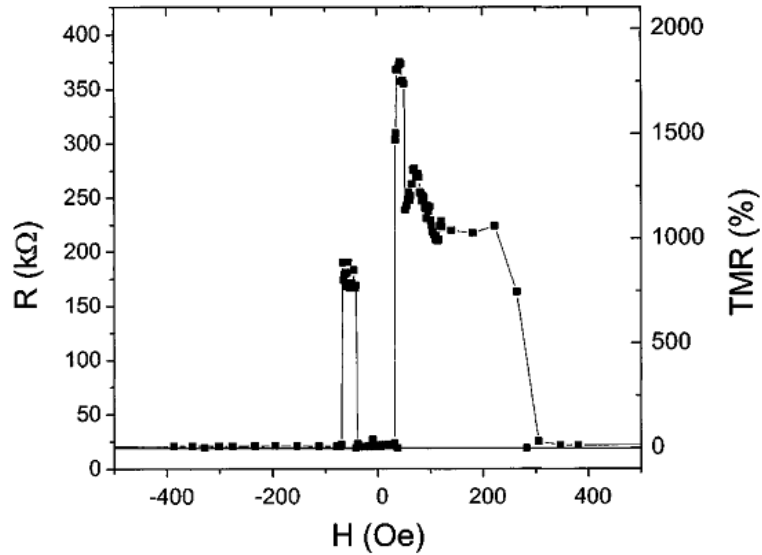

**Fig.20.** Magnetic field dependence of the junction resistance for LSMO/STO/LSMO/Co/Au device at 4.2 K.[97]

### 3.3.2 Four states and magnetoelectric coupling in MFTJ

Besides the TMR phenomena in MTJ mentioned above, the functionality of a tunnel junction can be enhanced by employing a ferroelectric material as the barrier layer. Recently, it has been theoretically[104-106] and experimentally[107-109] demonstrated that the tunneling conductance through a ferroelectric barrier depends on the polarization direction of the barrier, *i.e.*, a tunneling electroresistance (TER) effect. Therefore, combining a MTJ with a ferroelectric barrier, namely, a multiferroic tunnel junction (MFTJ) will add new functionalities to the MTJ and make the junction resistance be sensitive to not only the magnetization alignment of the ferromagnetic electrodes but also the polarization of the ferroelectric barrier.[110, 111] This makes a MFTJ device display a non-volatile four resistance states due to the coexistence of TMR and TER effects, and also a magnetoelectric device cross-controlled by magnetic and electric fields owing to the interfacial magnetoelectric coupling effect. Recently, such ferromagnetic/ferroelectric/ferromagnetic MFTJs with four non-volatile resistance states and magnetoelectric coupling were experimentally fabricated and investigated in Fe (or Co)/$BaTiO_3$/$La_{0.7}Sr_{0.3}MnO_3$,[112, 113] $La_{0.4}Sr_{0.3}MnO_3$/$BiFeO_3$/$La_{0.7}Sr_{0.3}MnO_3$,[114] and Co/$PbZr_{0.2}Ti_{0.8}O_3$/$La_{0.7}Sr_{0.3}MnO_3$[115] MFTJs.

Fig.21 shows the resistance versus magnetic field curves for a Co/PZT(3.2 nm)/LSMO junction which exhibits a TMR effect due to the ferromagnetic electrodes as well as a TER effect driven by the switching of the ferroelectric polarization of the barrier,[115] namely, the distinct four resistance states were obtained. In addition, it is noted that the switching of the ferroelectric polarization can also invert the sign of the spin polarization at the Co/PZT interface and lead to different sign of TMR at two polarization states. The magnetoelectric coupling at such a multiferroic interface could be due to hybridization at the interface, charge carrier doping, or spin-dependent screening in the magnetic electrodes.[115]

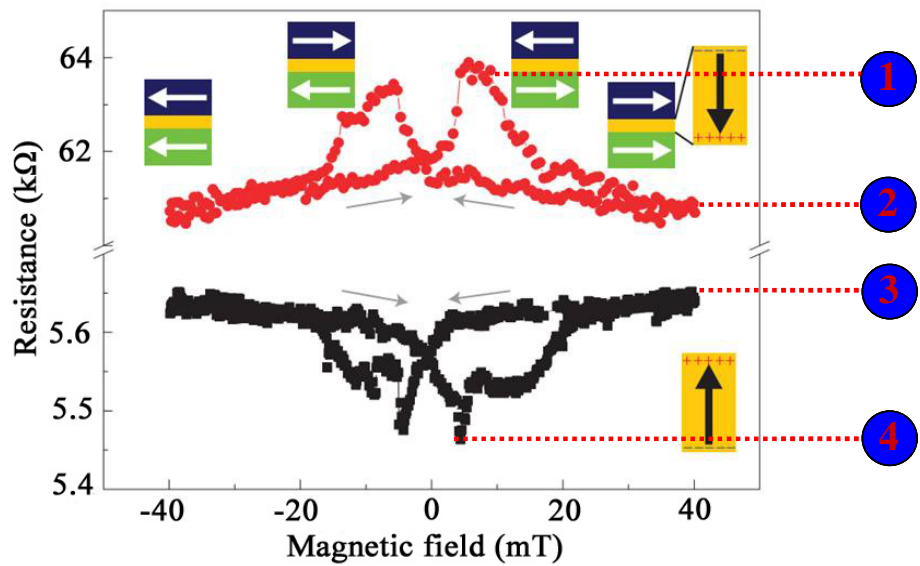

**Fig. 21.** Resistance versus magnetic field curves for a Co/PZT(3.2 nm)/LSMO junction measured at 50 K with the ferroelectric polarization pointing upward (black curve) and downward (red curve).[115]

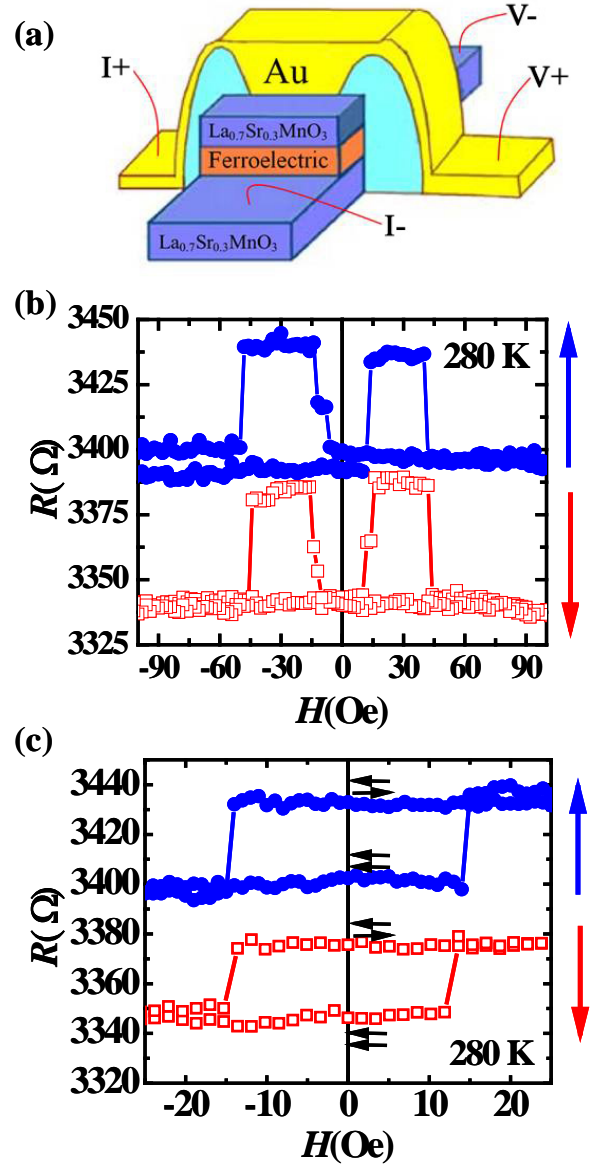

**Fig.22.** (a) Schematic diagram of the MFTJ device geometry. *R vs. H* loops between (a) ± 100 Oe and (b) ± 30 Oe for a La$_{0.7}$Sr$_{0.3}$MnO$_3$/Ba$_{0.95}$Sr$_{0.05}$TiO$_3$/La$_{0.7}$Sr$_{0.3}$MnO$_3$ MFTJ measured at 10 mV and 280 K after poling the ferroelectricity upward (red) and downward (blue). The horizontal arrows indicate the directions of the electrode magnetization and the vertical arrows indicate the directions of the barrier polarization.[116]

Using half-metallic manganite La$_{0.7}$Sr$_{0.3}$MnO$_3$ as ferromagnetic electrodes and nanometer thick (Ba, Sr)TiO$_3$ as ferroelectric barrier, Yin *et al.* have studied all-perovskite MFTJs as schematically shown in Fig.22(a), and observed four non-volatile states at room temperature, which demonstrates their application possibility.[116-118] As shown in Fig.22(b), the magnetic field dependence of the junction resistance for a La$_{0.7}$Sr$_{0.3}$MnO$_3$/Ba$_{0.95}$Sr$_{0.05}$TiO$_3$/La$_{0.7}$Sr$_{0.3}$MnO$_3$ MFTJ was measured after aligning the ferroelectric polarization of $Ba_{0.95}Sr_{0.05}TiO_3$ barrier downward (by +1.5 V poling voltage) or upward (by -1.5 V poling voltage).[116] It can be seen that the entire *R–H* curve shifts after reversing the polarization of barrier so that both parallel and antiparallel resistances switched to different values and four resistance states were observed. Standard MTJ memory loops can also be obtained for both polarization states and the resistance can be switched from one memory loop to another by reversing the ferroelectricity of the barrier, as shown in Fig.22(c). In other words, these non-volatile junction resistances between the two polarization states can be switched each other directly by reversing the barrier polarization repeatedly, as shown in Fig.23. Here, the red and blue dots mean that the switching was observed between parallel and antiparallel magnetic states [reached by *R–H* sweep (see Fig.23)], respectively. Hence, after applying positive and negative voltages, two parallel and two antiparallel states, *i.e.* a distinct four-state prototype device was represented at room temperature.

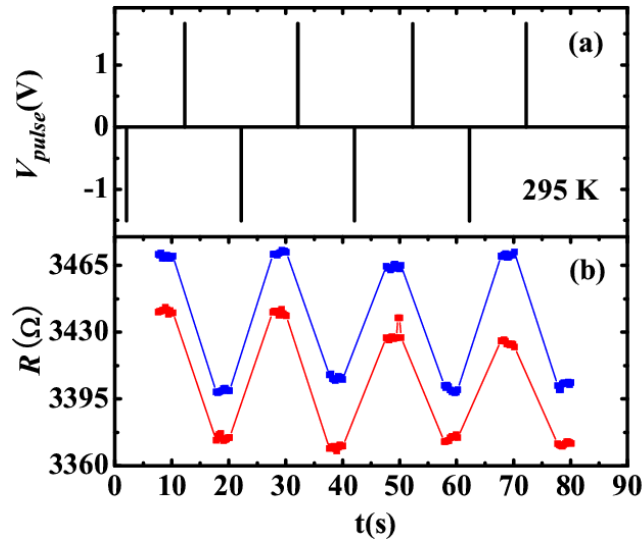

**Fig.23.** In response to (a) the applied voltage pulses, (b) switching of the junction resistance (measured at 10 μA) for a $La_{0.7}Sr_{0.3}MnO_3/Ba_{0.95}Sr_{0.05}TiO_3/La_{0.7}Sr_{0.3}MnO_3$ MFTJ at parallel (red) and antiparallel (blue) states.[116]

In a MFTJ, an interfacial magnetoelectric effect will show up at ferroelectric/ferromagnetic interface due to the chemical bonding, spin dependent screening, *etc*, therefore, the magnetization can be manipulated by electric field through a switchable ferroelectric polarization.[119, 120] Based on the intrinsic relationship between charge and magnetism in doped manganites $La_{1-x}A_xMnO_3$ (A =

Ca, Sr, or Ba) which exhibit rich phase diagrams as a function of hole concentration $x$,[7] the carrier density of $La_{1-x}A_xMnO_3$ near the interface with a ferroelectric material can be effectively altered by the screening charges and will result in another kind of ferroelectric control of magnetism in $La_{1-x}A_xMnO_3$.[31, 90, 121]

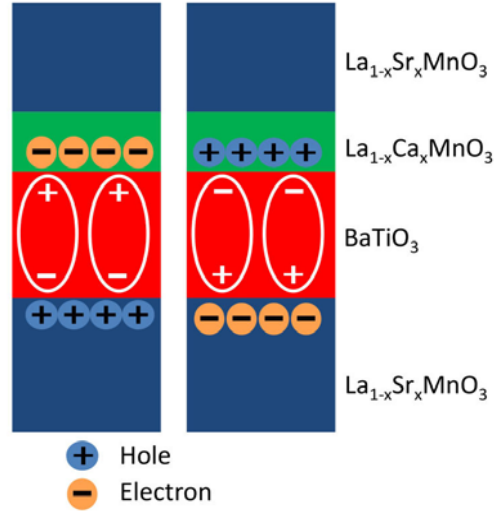

**Fig.24.** Schematic demonstration of the screening charge accumulation in electrodes for opposite ferroelectric polarization orientations in the BTO layer.[122]

Very recently, using first-principles density-functional calculations, Burton and Tsymbal[123] predicted a giant TER effect by exploiting a magnetoelectric interaction between a ferroelectric $BaTiO_3$ tunneling barrier and a magnetic $La_{1-x}Sr_xMnO_3$ electrode. According to this theory, Yin *et al* have designed and fabricated[122] the epitaxial $La_{0.7}Sr_{0.3}MnO_3/BaTiO_3/La_{0.5}Ca_{0.5}MnO_3/La_{0.7}Sr_{0.3}MnO_3$ (50 nm)/(3 nm)/(0.4-1.2 nm)/(30 nm) (bottom/barrier/top) tunnel junctions with a thin $La_{0.5}Ca_{0.5}MnO_3$ film inserted between the ferromagnetic $La_{0.7}Sr_{0.3}MnO_3$ electrode and ferroelectric $BaTiO_3$ barrier. On the one hand, for polarizing to the thin $La_{0.5}Ca_{0.5}MnO_3$ layer (see Fig.24), the screening electron accumulation or hole depletion will change the doping level of $La_{0.5}Ca_{0.5}MnO_3$ to $x < 0.5$ side which is in ferromagnetic metallic phase. While at the other side of $BaTiO_3$ barrier, because the stoichiometry of $La_{0.7}Sr_{0.3}MnO_3$ is far enough away from the phase boundary, the theoretical calculation demonstrated that the magnetic reconstruction will not occur.[124] On the other hand, for polarization pointing away from $La_{0.5}Ca_{0.5}MnO_3$ film, the electron depletion or hole accumulation will change $La_{0.5}Ca_{0.5}MnO_3$ to $x > 0.5$ side and push it into antiferromagnetic insulating phase. These a few unit-cells of antiferromagnetic $La_{0.5}Ca_{0.5}MnO_3$ may act as an atomic-scale spin valve by filtering spin-dependent current and increasing the junction resistance as well. In addition, the metallicity change of $La_{0.5}Ca_{0.5}MnO_3$ into insulation along [001] crystal direction will effectively increase the barrier thickness and reduce the tunneling conductance significantly. Thus, a much larger TER effect can be observed.

Fig.25(a) shows the resistance memory loops as a function of pulsed poling voltage for a MFTJ with 2 unit cells $La_{0.5}Ca_{0.5}MnO_3$ inserted, labeled as S1. Each datum point in the hysteresis curve is tested at a fixed voltage bias 10 mV after applying a voltage pulse. The resistance is higher when large enough positive voltage was applied to polarize the ferroelectricity downward, and will switch to a lower resistance state when ferroelectricity is reversed by large enough negative voltage bias. This is consistent with our design that the negative voltage will switch the ferroelectricity upward, pointing to the $La_{0.5}Ca_{0.5}MnO_3$ that will create a hole depletion as illustrated in Fig.24. This will push $La_{0.5}Ca_{0.5}MnO_3$ into a more ferromagnetic metallic phase. The TER ratio reaches up to ~5,000% and is much larger than the previous reported MFTJs in trilayer patterned junctions [112-114, 116, 117, 125]. For comparison, the $R$-$V_{pulse}$ loop of another junction without $La_{0.5}Ca_{0.5}MnO_3$, labeled as S2, was also measured and shown in Fig.25(b). The TER ratio for this junction is only ~30%, much smaller than S1.

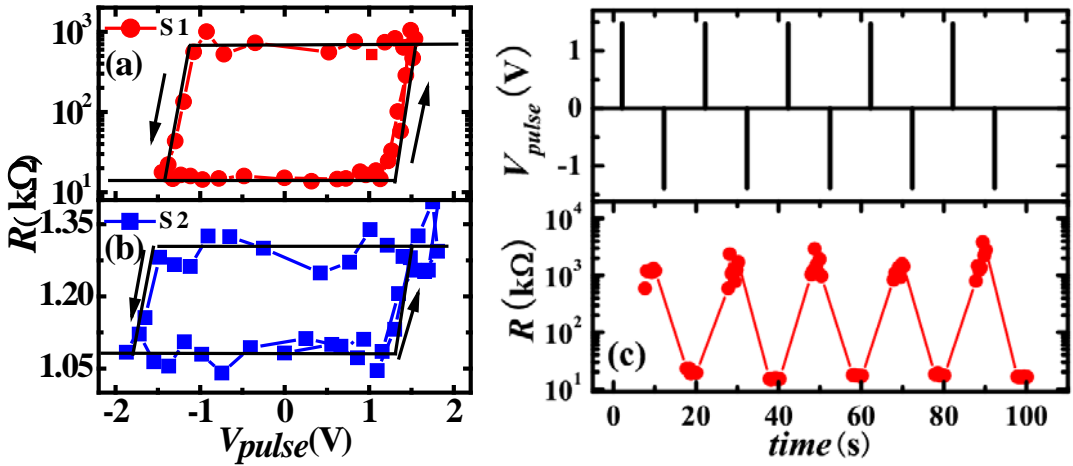

**Fig.25.** (a-b) Resistance memory loops as a function of pulsed poling voltage ($V_{pulse}$) at 40 K for MFTJs with (a) and without (b) $La_{0.5}Ca_{0.5}MnO_3$ inserted. The solid lines are a guide to the eyes. The arrows indicate the direction of pulse sequence. (c) Applied voltage pulse train (top panel) and the corresponding resistance switching (bottom panel) between positive and negative polarization states for S1 at 5 K and 10 mV.[122]

Besides the large magnitude of TER ratio, the reproducibility of polarization states is another decisive characteristic of a MFTJ for a memory device. The resistance states of S1 can be switched between positive and negative polarization states reproducibly. As shown in Fig.25(c), through a series of consecutive switching of the barrier polarization by voltage pulses (± 1.4 V), the resistance at 10 mV bias switches back and forth between two resistance states repeatedly and the TER value is as large as ~8,000%. The switching between two polarization states has been measured several hundred times at different temperatures and biases, and a large and reproducible TER effect can be observed accordingly, suggesting a possible nondestructive readout of polarization direction.[122]

### 3.3.3 Spin detector in organic spin-valves

Recently, organic semiconductors have attracted much attention in spintronics community, due to the long spin relaxation time. The so called organic spin-valve (OSV) is one of such devices based on organic semiconductor sandwiched between $La_{2/3}Sr_{1/3}MnO_3$ (LSMO) and metal electrodes. A larger spin-valve magnetoresistance effect discovered in the sandwich structures is associated with exotic properties of the magnetic electrode and organic semiconductors with long spin lifetimes.[98] It should be pointed out that although the organic spacers in OSVs are composed of light elements having weak spin-orbital interaction and consequently possessing long spin-relaxation times,[126] the hyperfine interaction (HFI) should have an important role in organic magnetotransport.[127] The role of the HFI in OSVs, by replacing hydrogen atoms ($^1$H, nuclear spin I=1/2) in the organic π-conjugated polymer poly (dioctyloxy) phenylenevinylene (DOO-PPV) spacer (dubbed here H-polymer) with deuterium atoms ($^2$H, I=1) (hereafter D-polymer) having much smaller HFI constant α, namely α(D)=α(H)/6.5, has been investigated with LSMO and Co as two ferromagnetic electrodes.[128] Fig.26 shows the magnetoresistance hysteresis loops for two similar OSVs with the organic layer thickness ~25 nm based on H- and D-polymers (devices a and b, respectively) at $T$=10 K and $V$=10 mV. It can be seen that the prototype devices based on the D-polymer show much larger MR values than those based on the H-polymer. Nguyen *et al* obtained that the spin-diffusion length $\lambda_s$ for D-poymers OSV is $\lambda_s$(D)=49 nm, whereas $\lambda_s$(H)=16 nm, which makes contribution to the improved magnetic properties for the D-polymer sandwiched OSVs.[128] The similar isotope effect on the magnetoresistance is also observed in Alq$_3$-based OSVs.[129] However, it should be noted that not all organic semiconductors would show a strong isotope effect,[130] presumably because interactions other than HFI may dominate the polaron-pair spin Hamiltonian [128] in the specific organic semiconductor, such as triplet-polaron interaction or/and spin orbit coupling.

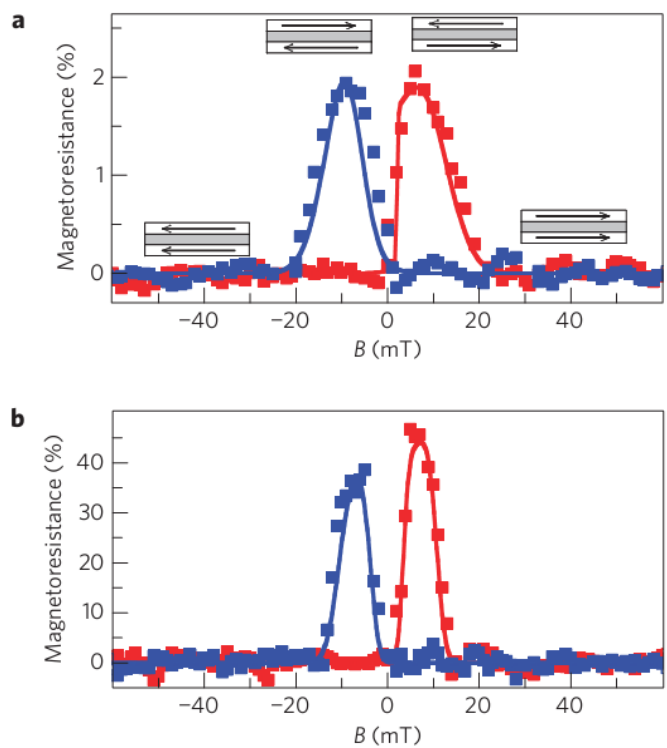

**Fig.26.** Isotope dependence of the magnetoresistance response in OSVs based on DOO-PPV polymers. Magnetoresistance loop of LSMO(200 nm)/DOO-PPV(25 nm)/Co(15 nm) spin-valve devices based on H- (a) and D- (b) polymers measured at $T$=10 K and $V$=10 mV, respectively. [128]

With the buckeyball $C_{60}$ molecule as the organic semiconductor spacer in OSVs, a 7% MR at $V$=100 mV and $T$=10 K is obtained, as shown in Fig.27(a).[131] Surprisingly, it is found that MR of $C_{60}$-based OSVs first increases as the interlayer $C_{60}$ thickness $d$ increases, reaches a maximum near $d_0$=35 nm, then exponentially decreases with $d$, as shown in Fig.27(b).[131] This is related to the $C_{60}$ film crystallinity which increases with $d$ below ~45 nm and saturates near 45 nm. The less disordered $C_{60}$ layer leads to higher charge carrier mobility and longer spin diffusion length $\lambda_{C_{60}}$ of $C_{60}$, thus results in a larger MR.[131] With further increasing $d$, more spins will be diffused and the MR reduces then.

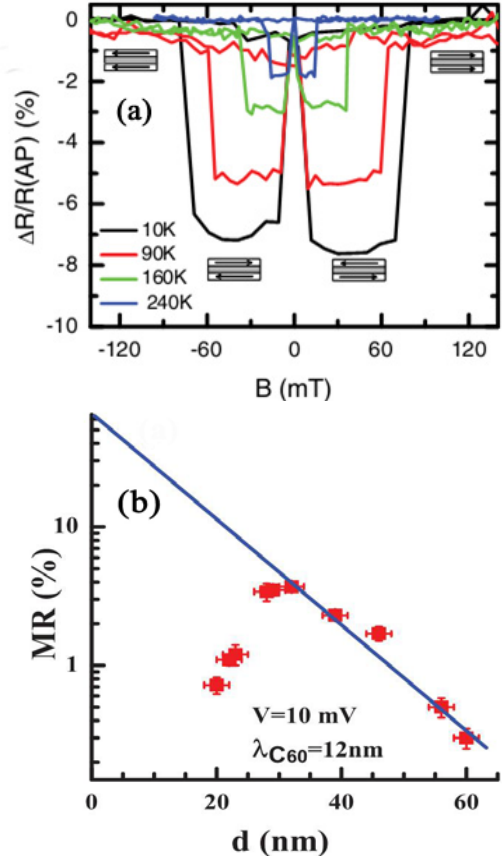

**Fig.27.** (a) MR of $C_{60}$-based OSVs measured at $V$=100 mV and different temperatures with $d$=35 nm. (b) Thickness of $C_{60}$ dependence of MR at $V$=10 mV and $T$=10 K.[131]

## 4. Nanosized manganite particles

### 4.1 Particle size effect on the magnetic properties of $La_{0.6}Pb_{0.4}MnO_3$

In the past few year, much attention has been drawn towards nanocrystalline CMR materials because when the size of these materials is reduced to nanometer scale, they exhibit a number of unique properties, such as low-field high magnetoresistance effect,[132, 133] superparamagnetism,[134] and surface spin-glass,[135, 136] as compared with their corresponding bulk materials.

As for the ferromagnetic manganites, the effect of the particles on the magnetic properties and its origin are controversial issues.[137-139] Mahesh *et al* reported that the $T_C$ of $La_{0.7}Ca_{0.3}MnO_3$ and $La_{0.7}Sr_{0.3}MnO_3$ nanoparticles decrease significantly with decreasing particle size.[137] However, Zheng *et al* found that the metal-insulator transition temperature $T_{Rmaxis}$ of the $La_{0.67}Ca_{0.33}MnO_3$ markedly decreases with decreasing the grain size, while the $T_C$ is not sensitive to the grain size.[138] While Dutta *et al* have found that the $T_C$ of $La_{0.875}Sr_{0.125}MnO_3$ nanoparticles increases with decreasing particle size and attributed this to domain status, changes in the Mn-O-Mn bond angle and Mn-O bond length, in comparison with bulk samples.[139, 140] The incensement of $T_C$ has also been found in $La_{0.67}Ca_{0.33}MnO_3$ nanoparticles with decreasing the particle size.[139, 140]

Recently, Zhang *et al* systemically studied the particle size effect on the structural and magnetic properties of ferromagnetic $La_{0.6}Pb_{0.4}MnO_3$ compounds with particle diameters $D$ varying from 5 to 100 nm.[141] It is found that the ferromagnetic transition temperature $T_C$ decreases clearly with decreasing $D$, as shown in Fig.28(a). The particle size dependence of $T_C$ can be described by an empirical formula as

$$T_C(D) = T_{C0}(1 - D_S / D)^\gamma \qquad (D > D_S), \tag{6}$$

where $T_{C0}$=360 K is the ferromagnetic transition temperature of the corresponding bulk material, $D_S$=5.8 nm is the upper particles limit for the appearance of superparamagnetism, and $\gamma$=0.68 is the fitting parameter. According to the double exchange model, $T_C$ is closely related to the magnitude of the O-2p-like bandwidth $W_{O_{2p}}$, and the particle size dependence of $W_{O_{2p}}$ can be expressed as

$$W_{O_{2p}} \propto \cos\omega / d_{Mn-O}^{3.5}, \tag{7}$$

where $\omega = (\pi - \theta_{Mn-O-Mn})/2$ is the tilt angle in the plane of the bond,[142] $\theta_{Mn-O-Mn}$ is the Mn-O-Mn bond angle, and $d_{Mn-O}$ is the Mn-O bond length which can be obtained from X-ray diffraction refinement.[141] From Fig.28(a), one can see that the relative value of $W_{O_{2p}}$ decreases with decreasing $D$. The decrease of $W_{O_{2p}}$ will result in the decrease of ferromagnetic transition temperature $T_C$, as depicted in the inset of Fig.28(a).

The broken exchange bonds and the increasing contribution of the surface effect with decreasing $D$ lead to a decrease of the saturated magnetization $M_S$, as shown in Fig.28(b).[141]. While with decreasing $D$ the magnetic coercive filed $H_C$ firstly increases, and reaches a maximum of 295 Oe for the 25 nm sample, then decreases. The variation of $H_C$ is closely related to the evolution of magnetic domain states with particle size. [143] For $D$>25 nm the $La_{0.6}Pb_{0.4}MnO_3$ nanoparticles have a multidomain structure; while for 5.8 nm<$D$≤25 nm the samples seem to have a single domain structure, and for the particle size less than 5.8 nm, the samples will show superparamagnetic behavior. Interestingly, for these superparamagnetic $La_{0.6}Pb_{0.4}MnO_3$ nanoparticles, they show memory effects in the *dc* magnetization and magnetic relaxation.[144]

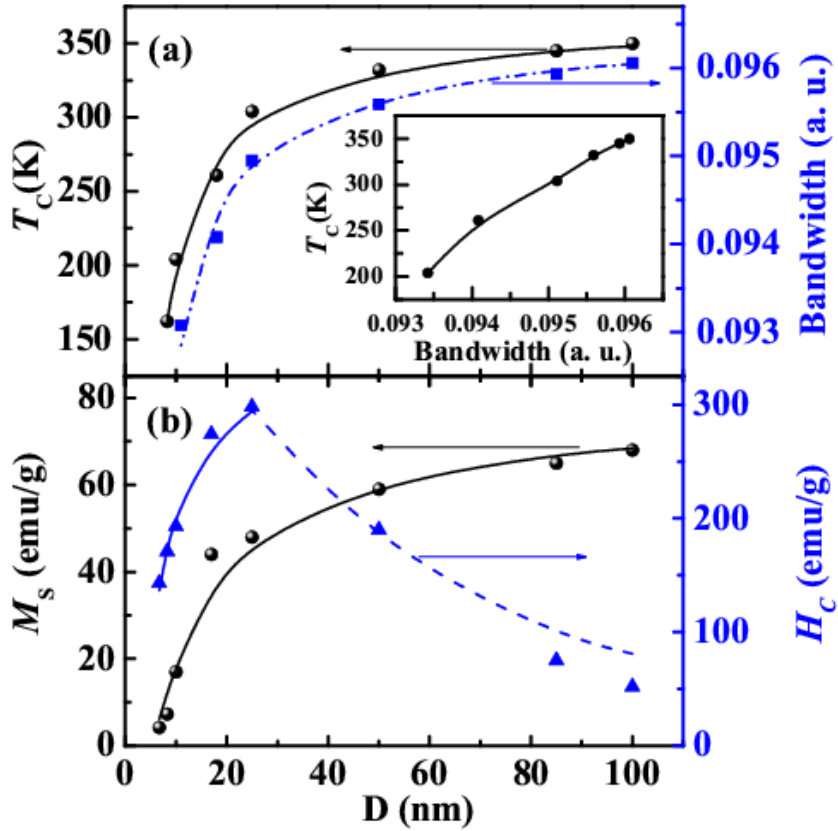

**Fig.28.** (a) Particle size dependences of ferromagnetic transition temperature $T_C$ and bandwidth $W_{O_{2p}}$ for $La_{0.6}Pb_{0.4}MnO_3$ nanoparticles. The solid line is the calculated result using Eq.(6) and the dash dot lone is obtained by the Eq.(7). Inset of (a) shows $W_{O_{2p}}$ dependence of $T_C$, the line is a guild for the eyes. (b) Variations of $M_S$ and $H_C$ as a function of particle size. [141]

## 4.2 Nanosized-induced CO melting and exchange bias effect in $La_{0.25}Ca_{0.75}MnO_3$ nanoparticles

For charge ordered manganites, it is believed that the stability of CO state is strong with the collinear AFM ordering of the localized Mn moments.[145] When the size is reduced to nanoscale, the uncompensated surface spins will destroy the long range collinear AFM configuration,[146] which, in turn, would impede the CO state. A series of experiments confirmed that with decreasing particle size there is a suppression of the CO states companied with a reduction of AFM phase as well as an occurrence of FM phase such as in the nanosized $La_{0.4}Ca_{0.6}MnO_3$,[147] $Nd_{0.5}Ca_{0.5}MnO_3$,[148] and $Pr_{0.5}Ca_{0.5}MnO_3$ systems,[149] *etc*.

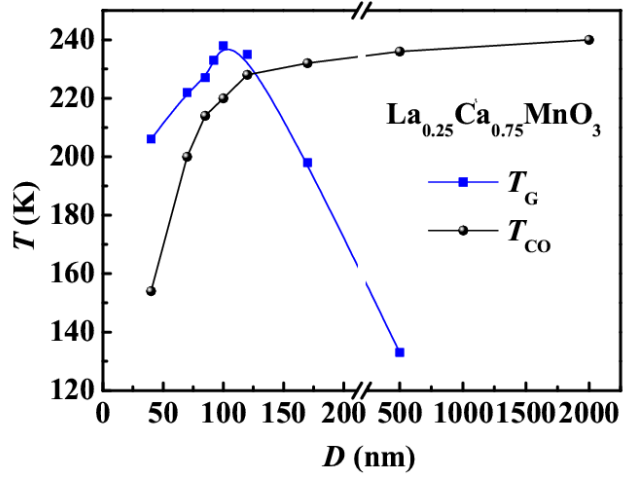

**Fig.29.** Temperature *vs.* particle size phase diagram of $La_{0.25}Ca_{0.75}MnO_3$. The solid lines are guide for the eyes.[150]

It is know that the CO state in the $La_{0.25}Ca_{0.75}MnO_3$ bulk compound is very stable, see Fig.9.[61] However, for the nanosized $La_{0.25}Ca_{0.75}MnO_3$, the CO transition gradually shifts to lower temperature with decreasing the particles size ($D$); meanwhile, the ferromagnetic cluster glass state appears,[151] and the spin cluster glass temperature $T_G$ shows a nonmonotonous variation with $D$ as shown in Fig.29.[150, 152] The variations of $T_{CO}$ and $T_G$ with particle size can be explained by the improved core-shell model.[150]. As illustrated in Fig.30, a bulk $La_{0.25}Ca_{0.75}MnO_3$ is of almost a perfect AFM spin ordering, while for a $La_{0.25}Ca_{0.75}MnO_3$ nanoparticle, a larger number of uncompensated surface spins appear, and the shell spins will deviate from an AFM arrangement of the core spins, which weakens the AFM interaction across the shell and hence enhances the FM interaction.[153] The enhanced FM coupling facilitates the formation of FM cluster glass in these regions, and thus the temperature $T_G$ for the appearance of ferromagnetic clusters increases with decreasing particles size. However, with further decreasing particle size, the surface spins become more disordered and thus disfavor the FM coupling, resulting in a decrease of $T_G$ with the decrease of $D$.[141, 148] Thus, when the two competitive factors reach a balance, $T_G$ attains a maximum. The destruction of AFM configuration, on the other hand, will cause a reduction of $T_{CO}$ with decreasing the particle size.[150] Furthermore, from the frequency and *dc* magnetic field dependencies of the *ac* susceptibilities of $La_{0.25}Ca_{0.75}MnO_3$ nanoparticles, one can find that the FM cluster size in the shell of the nanoparticles increases with decreasing particle size in a certain size range.[136]

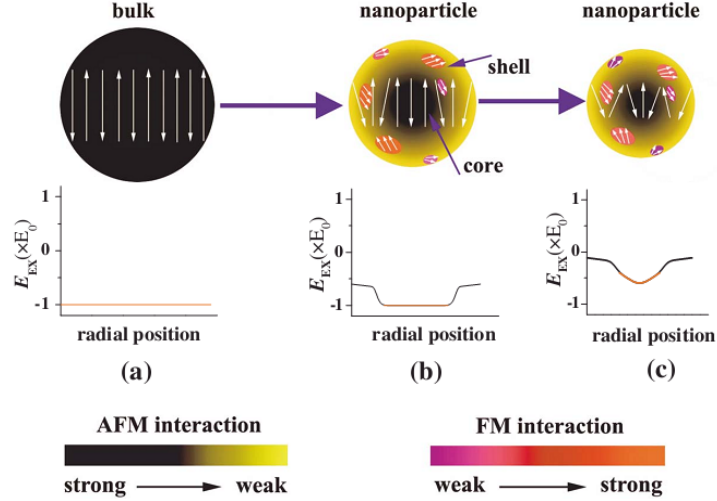

**Fig.30.** Schematic diagrams (two dimensional) for evolutions of the magnetic configuration, the exchange energy, and magnetic interactions with particle size, proposed on the basis of the core-shell model in Ref. 156. Arrows only represent the spin orientations. (a) Perfect AFM spin ordering in a bulk $La_{0.25}Ca_{0.75}MnO_3$, assuming the AFM exchange energy $E_{ex}=-E_0$. (b) For nanoparticles, the core spins are nearly antiparallel, the deviation of shell spins from AFM arrangement makes the $E_{ex}$ value less negative and leads to the formation of FM cluster across the shell regions. (c) For smaller nanoparticles, more disordered shell spins not only induce the deviation of core spins from AFM arrangement and accordingly further increase $E_{ex}$ value but also weaken the FM coupling greatly.[150]

Because of the coexistence of FM cluster and AFM phase, the exchange bias (EB) effect should be observed in those manganite nanoparticles.[154, 155] Fig.31 shows the particle diameter $D$ dependencies of exchange bias field ($H_{EB}$) and coercivity ($H_C$) for $La_{0.25}Ca_{0.75}MnO_3$ at different temperatures.[154]. It can be seen that with increasing temperature, $H_{EB}$ decreases and the peak height diminishes gradually but the peak position (around 80 nm) changes inconspicuously. It is believed that the exchange coupling is closely related to the interfacial uncompensated spins, FM/AFM interfacial pinning strength as well as the AFM anisotropy energy. Thus, the diameter dependencies of $H_{EB}$ can be well explained in terms of the core-shell model in Fig.30. However, the peak position of $H_C$ shifts to a larger particle size with increasing temperature, inconsistent with that of $H_{EB}$.

From the results mentioned above, one can see that there are some basic topics in the manipulating physical properties which need further investigations, although the distinct physical and chemical properties in nanosized CMR manganites show potential applications in mesoscopic physics and nanodevices.[156-158]

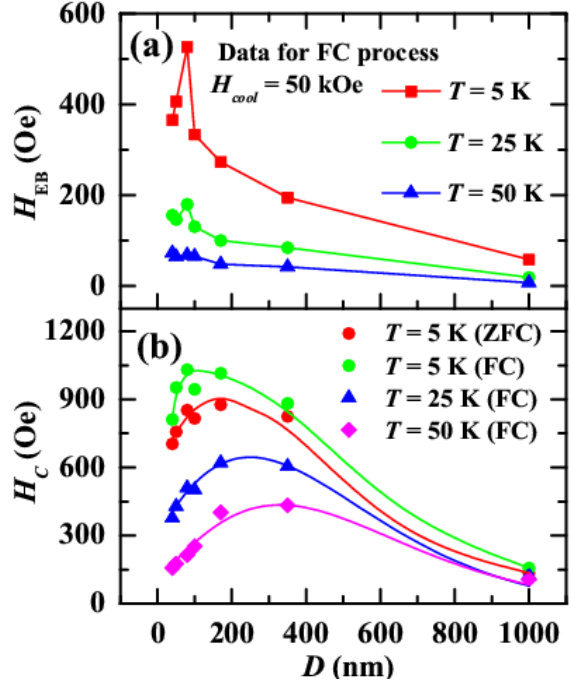

**Fig.31.** Particle diameter dependencies of (a) exchange bias field $H_{EB}$ measured at 5, 25, and 50 K in FC processes for $La_{0.25}Ca_{0.75}MnO_3$, and (b) coercivity $H_C$ measured at 5 K in ZFC and FC processes and at 25 and 50 K in FC processes. The solid lines are guides for the eyes.[154]

## 5. Conclusion and remarks:

In summary, the colossal magnetoresistance effects in manganites related to the metal-insulator transition and charge ordered state melting in magnetic fields were discussed. A phenomenological model describing the relationship between phase separation and magnetoresistance is proposed, which shows that the competition between the metallic and insulating phases plays an important role in the CMR effect. Based on the CMR effect and high spin-polarization in manganites, some prototype devices have been fabricated and show potential applications in multifunctional spintronics. In the manganite/Nb:STO *p-n* junctions, electric field and temperature tuned positive/negative magnetoresistance effect due to the band structure change at the manganites-Nb:STO interface has been discussed. Using ferromagnetic manganites as metallic electrodes in multiferroic heterostructures, one can realize ferroelectric and magnetic field controlled electric and magnetic properties of manganites as well as the manganite/ferroelectric heterostructures. Using half-metallic property of manganites, the high TMR effect in magnetic tunnel junction, four nonvolatile states and magnetoelectric coupling in multiferroic tunnel junction, and spin detector in organic spin-valve are presented. In addition, the effect of nanoparticle size on the FM $La_{0.6}Pb_{0.4}MnO_3$ nanoparticles and magnetic properties in CO $La_{0.25}Ca_{0.75}MnO_3$ system are discussed.

The CMR manganites show rich physics with complex electronic, magnetic, and structural phase diagrams and great potential application in multifunctional spintronics, however, from the industrial viewpoint, some obstacles have to be overcome: (*i*) How to integrate the new phenomena to CMR manganites, such as the magnetoelectric coupling, spin dependent tunneling, resistance switching, spin transfer torque, *etc*., to realize low-field high magnetoresistance effect, new functionality and high memory density. (*ii*) How to incorporate the prototype devices based on colossal magnetoresistance onto a semiconductor chip, while not sacrificing the emergent phenomena in manganite. (*iii*) Compared to the bulk, the high quality films show much more powerful application prospect, such as in tunnel junctions and spin-valves. However, the high growth temperature (usually 700℃ or higher) for preparing the high quality thin films by sputtering, molecular epitaxy, pulsed laser deposition, *etc.*, requiring high energy cost, is not currently used in conventional process of microelectronics.

In allusion to these questions, a considerable research has been done. For example, low-temperature synthesis methods, such as sol-gel technique,[159, 160] chemical solution deposition,[161, 162] wet-chemical method,[163] *etc*., have been developed to fabricate CMR manganite film on single crystals. In order to integrate perovskite-type manganites onto silicon, different templates or buffer layers between the CMR layer and silicon, such as $Bi_4Ti_3O_{12}$,[164, 165] $SrTiO_3$,[166-168] yttrium-stabilized zirconia (YSZ),[169, 170] have been used. Despite all these studies, the problems mentioned above are still not solved well, and further studies are needed for exploring new materials and new technologies.

**Acknowledgements**

This work is supported by the National Natural Science Foundation of China and the National Basic Research Program of China (Contract Nos. 2012CB922003, 2011CBA00102, and 2009CB929502).